\documentclass[conference]{IEEEtran}
\ifCLASSINFOpdf
\else
\fi
\usepackage[cmex10]{amsmath}
\usepackage{algorithm}
\usepackage{algorithmic}
\usepackage[tight,footnotesize]{subfigure}
\usepackage{graphicx}
\usepackage{color}

\begin{document}
\bstctlcite{IEEEexample:BSTcontrol}
%
% paper title
% can use linebreaks \\ within to get better formatting as desired
\title{Parallel Toolkit for Measuring the Quality of Network Community Structure}

% author names and affiliations
% use a multiple column layout for up to three different
% affiliations
\author{\IEEEauthorblockN{Mingming Chen}
\IEEEauthorblockA{Department of Computer Science\\
Rensselaer Polytechnic Institute\\
110 8th Street, Troy, NY 12180\\
Email: chenm8@rpi.edu}
\and
\IEEEauthorblockN{Sisi Liu}
\IEEEauthorblockA{Information Technology and Web Science\\
Rensselaer Polytechnic Institute\\
110 8th Street, Troy, NY 12180\\
Email: lius10@rpi.edu}
\and
\IEEEauthorblockN{Boleslaw K. Szymanski}
\IEEEauthorblockA{Department of Computer Science\\
Rensselaer Polytechnic Institute\\
110 8th Street, Troy, NY 12180\\
Email: szymab@rpi.edu}
}

% make the title area
\maketitle

\begin{abstract}
%\boldmath
Many networks display community structure which identifies groups of nodes within which connections are denser than between them. Detecting and characterizing such community structure, which is known as community detection, is one of the fundamental issues in the study of network systems. It has received
a considerable attention in the last years. Numerous techniques have been developed for both efficient and effective community detection. Among them, the most efficient algorithm is the label propagation algorithm whose computational complexity is $O(|E|)$. Although it is linear in the number of edges, the running time is still too long for very large networks, creating the need for parallel community detection. Also, computing community quality metrics for community structure is computationally expensive both with and without ground truth. However, to date we are not aware of any effort to introduce parallelism for this problem. In this paper, we provide a parallel toolkit\footnote{Please contact Mingming Chen via mileschen2008@gmail.com for the parallel toolkit if you are interested in it.} to calculate the values of such metrics. We evaluate the parallel algorithms on both distributed memory machine and shared memory machine. The experimental results show that they yield a significant performance gain over sequential execution in terms of total running time, speedup, and efficiency.
\end{abstract}

% For peerreview papers, this IEEEtran command inserts a page break and
% creates the second title. It will be ignored for other modes.
\IEEEpeerreviewmaketitle

\section{Introduction}
Many networks, including Internet, citation networks, transportation networks, e-mail networks, and social and biochemical networks, display community structure which identifies groups of nodes within which connections are denser than between them \cite{UWDModularity}. Detecting and characterizing such community structure, which is known as community detection, is one of the fundamental issues in the study of network systems. Community detection has been shown to reveal latent yet meaningful structures in networks \cite{CommunityReport}. % such as groups in online and contact-based social networks, functional modules in protein-protein interaction networks, groups of customers with similar interests in online retailer user networks, groups of scientists in interdisciplinary collaboration networks, etc. \cite{CommunityReport}.

Thus, numerous techniques were developed for both efficient and effective community detection, including Modularity Optimization \cite{NewmanGreedy,PNASModularity}, Clique Percolation \cite{CPM}, Local Expansion \cite{LocalExpansionRPI,LFM}, Fuzzy Clustering \cite{ZhangFuzzy,NMFFuzzy}, Link Partitioning \cite{LinkPartition}, and Label Propagation \cite{LPA,SLPA2012,LabelRankT}. Among them, the most efficient algorithm is the label propagation algorithm whose computational complexity is $O(|E|)$, where $|E|$ is the number of edges in the network. Although it is a linear algorithm, the running time is still too long for very large networks. The primary examples are online social networks that are increasingly popular, the largest being Facebook with more than 800 million daily active users\footnote{Facebook company info: http://newsroom.fb.com/company-info/}. The WWW forms a network of hyperlinked webpages in excess of 30 billion nodes. Therefore, parallelism was introduced into community detection to alleviate computational costs \cite{ParallelPropinquity,ParallelMassive,ParallelKClique,ParallelSLPA,TwoParallelAlg}. However, to date we are not aware of any effort that provides parallel computation for the community quality metrics with and without ground truth community structure \cite{fine_tuned,QdsConference,QdsJournal}, though it is computational expensive to do so. Hence, in this paper, we provide a parallel toolkit to calculate the values of these metrics. Although we are using parallel computing to speed up the processing, in most of the cases, algorithms are highly parallelizable, so the contributions of this paper focus on making the highly efficient social network analysis tools available to research community.

We implement the parallel algorithms with MPI (Message Passing Interface) and Pthreads (POSIX Threads). We perform runs on both distributed memory machine, such as Blue Gene/Q, and shared memory machine, like GANXIS. The network we adopt is LFR benchmark network \cite{LFR}. The LFR benchmark network for testing the parallel programs to calculate the metrics with ground truth community structure has $100,000$ nodes. We choose two sizes, ten million of nodes ($10,000,000$) and one hundred million of nodes ($100,000,000$), of the LFR benchmark network to test the parallel programs for computing the metrics without ground truth communities. The experimental results show that both parallel MPI algorithms and parallel Pthreads algorithms yield a significant performance gain over sequential execution. Moreover, we recommend using parallel MPI algorithms and parallel Pthreads algorithm respectively to calculate the metrics with and without ground truth communities on GANXIS (or shared memory machines) in terms of their speedup and efficiency.

\section{Community Quality Metrics}
\label{sec:metric}
%In this section, we will introduce the definitions of the community quality metrics both with and without ground truth community structure.
\subsection{Metrics with Ground Truth Communities}
\vspace{-0.2em}
\label{sec:CQM_with}
The quality evaluation metrics with ground truth community structure we consider here can be divided into three categories: \textit{Variation of Information} ($VI$) and \textit{Normalized Mutual Information} ($NMI$) based on information theory; \textit{F-measure} and \textit{Normalized Van Dongen metric} ($NVD$) based on cluster matching; \textit{Rand Index} ($RI$), \textit{Adjusted Rand Index} ($ARI$), and \textit{Jaccard Index} ($JI$) based on pair counting \cite{fine_tuned}.

\subsubsection{Information Theory Based Metrics}
Given partitions $C$ and $C'$, $VI$ quantifies the ``distance'' between those two partitions, while $NMI$ measures the similarity between them. $VI$ is defined as
\begin{equation}
\label{eq:vi}
VI(C,C')=-\frac{1}{|V|}\sum_{c \in C, c' \in C'} |c \cap c'| \log \left(\frac{|c \cap c'|^2}{|c||c'|}\right),
\end{equation}
and $NMI$ is given by
\begin{equation}
\label{eq:nmi}
NMI(C,C')=\frac{-2\sum_{c \in C, c' \in C'} \frac{|c \cap c'|}{|V|} \log \left(\frac{|c \cap c'||V|}{|c||c'|}\right)}{\sum_{c \in C}\frac{|c|}{|V|}\log \frac{|c|}{|V|}+\sum_{c' \in C'}\frac{|c'|}{|V|}\log \frac{|c'|}{|V|}},
\end{equation}
where $|V|$ is the number of nodes in the network, $|c|$ is the number of nodes in community $c$ of $C$, and $|c \cap c'|$ is the number of nodes both in community $c$ of $C$ and in community $c'$ of $C'$. The computational complexity to calculate $VI$ and $NMI$ is $O(|V||C'|)$, where $|C'|$ is the number of communities found by a community detection algorithm.

\subsubsection{Cluster Matching Based Metrics}
Measures based on cluster matching aim at finding the largest overlaps between pairs of communities of two partitions $C$ and $C'$. \textit{F-measure} measures the similarity between two partitions, while $NVD$ quantifies the ``distance'' between partitions $C$ and $C'$. \textit{F-measure} is defined as
\begin{equation}
\label{eq:fmeasure}
F\textit{-}measure(C,C')=\frac{1}{|V|}\sum_{c \in C} |c| \max_{c' \in C'} \frac{2|c \cap c'|}{|c|+|c'|}.
\end{equation}
$NVD$ is given by
\begin{equation}
\label{eq:nvd}
NVD(C,C')=1-\frac{1}{2|V|} \biggl(\sum_{c \in C} \max_{c' \in C'} |c \cap c'|+\sum_{c' \in C'} \max_{c \in C} |c' \cap c| \biggr).
\end{equation}
%\begin{equation}
%\label{eq:nvd}
%\begin{split}
%&NVD(C,C')=1-\frac{1}{2|V|} \biggl(\sum_{c \in C} \max_{c' \in C'} |c \cap c'|  \\
%&~~~~~~~~~~~~~~~~~~~+\sum_{c' \in C'} \max_{c \in C} |c' \cap c| \biggr).
%\end{split}
%\end{equation}
The complexity to calculate \textit{F-measure} and $NVD$ is $O(|V|(|C|+|C'|))$, where $|C|$ is the number of communities in the ground truth community structure.

\subsubsection{Pair Counting Based Metrics}
Metrics based on pair counting count the number of pairs of nodes that are classified (in the same community or in different communities) in two partitions $C$ and $C'$. Let $a_{11}$ indicate the number of pairs of nodes that are in the same community in both partitions, $a_{10}$ denote the number of pairs of nodes that are in the same community in $C$ but in different communities in $C'$, $a_{01}$ be the number of pairs of nodes which are in different communities in $C$ but in the same community in $C'$, $a_{00}$ be the number of pairs of nodes which are in different communities in both partitions. By definition, $A=a_{11}+a_{10}+a_{01}+a_{00}=\frac{|V|(|V|-1)}{2}$ is the total number of pairs of nodes in the network. Then, $RI$ which is the ratio of the number of node pairs placed in the same way in both partitions to the total number of pairs is given by
\begin{equation}
\label{eq:ri}
RI(C,C')=\frac{a_{11}+a_{00}}{A}.
\end{equation}
Denote $M=\frac{1}{A}(a_{11}+a_{10})(a_{11}+a_{01})$. Then, \textit{RI}'s corresponding adjusted version, $ARI$, is expressed as
\begin{equation}
\label{eq:ari}
ARI(C,C')=\frac{a_{11}-M}{\frac{1}{2}\left[(a_{11}+a_{10})+(a_{11}+a_{01})\right]-M}.
\end{equation}
$JI$ which is the ratio of the number of node pairs placed in the same community in both partitions to the number of node pairs that are placed in the same group in at least one partition is defined as
\begin{equation}
\label{eq:ji}
JI(C,C')=\frac{a_{11}}{a_{11}+a_{10}+a_{01}}.
\end{equation}
Each of these three metrics quantifies the similarity between two partitions $C$ and $C'$. The complexity to calculate $RI$, $ARI$, and $JI$ is $O(|V|^2)$.

\subsection{Metrics without Ground Truth Communities}
\label{sec:CQM_without}
\vspace{-0.3em}
\subsubsection{Newman's Modularity}
Modularity \cite{UWDModularity} measures the difference between the actual fraction of edges within the community and such fraction expected in a randomized graph with the same number of nodes and the same degree sequence. For the given community partition of a unweighted and undirected graph $G=(V,E)$ with $|E|$ edges, modularity ($Q$) is given by
\begin{equation}
\label{eq:uwdmodularity}
Q=\sum_{c \in C} \left[\frac{|E_{c}^{in}|}{|E|}-\left(\frac{2|E_{c}^{in}|+|E_{c}^{out}|}{2|E|}\right)^2\right],
\end{equation}
where $C$ is the set of all the communities, $c$ is a specific community in $C$, $|E_{c}^{in}|$ is the number of edges between nodes within community $c$, and $|E_{c}^{out}|$ is the number of edges from the nodes in community $c$ to the nodes outside $c$.

\subsubsection{Modularity Density}
\textit{Modularity Density} ($Q_{ds}$) \cite{QdsConference,QdsJournal} is proposed to solve the two opposite yet coexisting problems of modularity: in some cases, it tends to favor small communities over large ones while in others, large communities over small ones. The latter tendency is also known as the resolution limit problem \cite{ResolutionLimit}. For unweighted and undirected networks, $Q_{ds}$ is defined as
\begin{equation}
\label{eq:uwdqds}
\begin{split}
&Q_{ds}=\sum_{c_i \in C} \biggl[\frac{|E_{c_i}^{in}|}{|E|}d_{c_i}-\left(\frac{2|E_{c_i}^{in}|+|E_{c_i}^{out}|}{2|E|}d_{c_i}\right)^2 \\
& ~~~~~~~~~~~~~-\sum_{\substack{c_j \in C \\ c_j \ne c_i}}\frac{|E_{c_i,c_j}|}{2|E|}d_{c_i,c_j}\biggr], \\
& d_{c_i}=\frac{2|E_{c_i}^{in}|}{|c_i|(|c_i|-1)}, \\
& d_{c_i,c_j}=\frac{|E_{c_i,c_j}|}{|c_i||c_j|}.
\end{split}
\end{equation}
In the above, $d_{c_i}$ is the internal density of community $c_i$, $|E_{c_i,c_j}|$ is the number of edges from $c_i$ to $c_j$, and $d_{c_i,c_j}$ is the pair-wise density between communities $c_i$ and $c_j$.
%Note that $|E_{c_i}^{in}|$ in $d_{c_i}$ and $|E_{c_i,c_j}|$ in $d_{c_i,c_j}$ are unweighted for both unweighted and weighted networks, so that those two community densities are always less than or equal to 1.0.

\subsubsection{Six Other Community Quality Measures}
\label{subsec:metrics}
We also consider six other metrics without ground truth community structure, including the number of \textit{Intra-edges}, \textit{Intra-density}, \textit{Contraction}, the number of \textit{Inter-edges}, \textit{Expansion}, and \textit{Conductance} \cite{QdsConference,QdsJournal}, which characterize how community-like is the connectivity structure of a given set of nodes. All of them rely on the intuition that communities are sets of nodes with many edges inside and few edges outside. \\
\textbf{The number of \textit{Intra-edges}:} $|E_c^{in}|$; it is the total number of edges in community $c$. \\ %A large value of this metric is better than a small value in terms of the community quality. \\
\textbf{\textit{Intra-density}:} $d_{c_i}$ in Equation~(\ref{eq:uwdqds}). \\ %The larger the value of this metric, the higher the quality of the communities. \\
\textbf{\textit{Contraction}:} $2|E_c^{in}|/|c|$; it measures the average number of edges per node inside the community $c$. \\ %A larger value of contraction means better community quality. \\
\textbf{The number of \textit{Inter-edges}:} $|E_c^{out}|$; it is the total number of edges on the boundary of $c$. \\ %A small value of this metric is better than a large value in terms of the community quality.  \\
\textbf{\textit{Expansion}:} $|E_c^{out}|/|c|$; it measures the average number of edges (per node) that point outside the community $c$. \\ %A smaller value of expansion corresponds to better community structure. \\
\textbf{\textit{Conductance}:} $\frac{|E_c^{out}|}{2|E_c^{in}|+|E_c^{out}|}$; it measures the fraction of the total number of edges that point outside the community. \\ %A smaller value of conductance means better quality of community structure.

\section{Parallel Algorithm Design}
\vspace{-0.4em}
In this section, we present the parallel algorithms, MPI and Pthreads versions, to calculate the quality metrics introduced in Section~\ref{sec:metric}. It can be seen from Section~\ref{sec:metric} that $VI$ and $NMI$ can be calculated together, \textit{F-measure} and $NVD$ can be computed together, $RI$, $ARI$, and $JI$ can be calculated together, and the metrics without ground truth communities can be computed together. Thus, we will have four parallel algorithms based on MPI and four parallel algorithms based on Pthreads for these metrics. We denote the ground truth community structure as $C$ and the community structure detected with a community detection algorithm as $C'$.

\vspace{-0.2em}
\subsection{Parallel Algorithms Based on MPI}
\label{sec:parallel_mpi}
\vspace{-0.4em}
In the parallel algorithms based on MPI, the problem to calculate metrics is partitioned with the unit of community. For the algorithms to calculate the metrics with ground truth community structure, each processor extracts the ground truth communities and the detected communities when ($comId\mod{numProcs}==procId$) to achieve rough load balance. $comId$ is the id of a community, $procId$ is the id of a processor, and $numProcs$ is the total number of processors used. Hence, each processor will have $|C|/numProcs$ or $|C|/numProcs+1$ ground truth communities and $|C'|/numProcs$ or $|C'|/numProcs+1$ discovered communities. For the algorithms to compute the metrics without ground truth communities, each processor extracts the discovered communities using the same approach. Also, each processor gets its local network which contains the nodes in its own communities and the neighboring nodes of these nodes.

\begin{algorithm}
\caption{MPI\_information\_theory\_metric($C_p$, $C'_p$)}
\label{alg:entropy_mpi}
\begin{algorithmic}[1]
  \STATE // Processor id is denoted as $p$. The number of processors used is denoted as $numProcs$. The local values of $VI$ and $NMI$ is denoted as $rankVI$ and $rankNMI$.
  %\STATE // Read its own ground truth communities.
%  \FOR{$i=0$ to $|C|-1$}
%    \IF{ $i\mod{numProcs}==p$}
%      \STATE $c$ = $C$.get($i$);
%      \STATE $C_p$.add($c$);
%    \ENDIF
%  \ENDFOR
%  \STATE // Read its own discovered communities.
%  \FOR{$i=0$ to $|C'|-1$}
%    \IF{ $i\mod{numProcs}==p$}
%      \STATE $c'$ = $C'$.get($i$);
%      \STATE $C'_p$.add($c'$);
%    \ENDIF
%  \ENDFOR
  \STATE Calculate $rankVI$ and $rankNMI$ based on Equation~(\ref{eq:vi}) and Equation~(\ref{eq:nmi}) using its own $C_p$ and its own $C'_p$;
  \STATE // Circulate $C'_p$ in a ring.
  \STATE $recvSrc=(p+numProcs-1)\mod{numProcs}$;
  \STATE $sendDst=(p+1)\mod{numProcs}$;
  \STATE $receivedMsgNum=0$;
  \WHILE{$receivedMsgNum<(numProcs-1)$}
    \STATE $sendBuf[] \leftarrow C'_p$;
    \STATE $\text{Send}~sendBuf[]~\text{to}~sendDst$;
    \STATE $\text{Receive}~C'_p~\text{from}~recvSrc$ with $recvBuf[]$;
    \STATE Calculate $rankVI$ and $rankNMI$ based on Equations~(\ref{eq:vi}) and~(\ref{eq:nmi}) using its own $C_p$ and received $C'_p$;
    \STATE ++$receivedMsgNum$;
  \ENDWHILE
  \STATE Get $VI$ and $NMI$ by summing the $rankVI$ and $rankNMI$ of all $numProcs$ processors;
  \STATE Return $VI$ and $NMI$;
\end{algorithmic}
\end{algorithm}

We first show the parallel MPI algorithm to calculate the information theory based metrics, $VI$ and $NMI$. Supposed there are $N$ processors (or MPI ranks), each processor reads its own set of ground truth communities $C_p$ and its own set of discovered communities $C'_p$. It can be learnt from the definitions of $VI$ and $NMI$ given by Equation~(\ref{eq:vi}) and Equation~(\ref{eq:nmi}) respectively that each ground truth community should traverse all the discovered communities in order to get the values of them. Thus, in the algorithm, we circulate $C'_p$ to each processor in a ``ring''. That is, processor 0 sends its $C'_p$ to processor 1, and processor 1 to processor 2, and processor $N-1$ would send $C'_p$ to processor 0. This ``shifting'' of $C'_p$ will occur $N-1$ times for $N$ processors. For its own $C'_p$ and each received $C'_p$, the processor will calculate its local values of $VI$ and $NMI$ with its own $C_p$. Finally, the values of $VI$ and $NMI$ are the sum of the local values of all $N$ processors. Algorithm~\ref{alg:entropy_mpi} shows our parallel algorithm for computing $VI$ and $NMI$ based on MPI. It takes $C_p$ and $C'_p$ as parameters. %This algorithm does not change the complexity of calculation, thus the complexity is still $O(|V||C'|)$.

\begin{algorithm}
\caption{MPI\_cluster\_matching\_metric($C_p$, $C'_p$)}
\label{alg:clustering_mpi}
\begin{algorithmic}[1]
  %\STATE // Read its own ground truth communities.
  %\STATE Read $C_p$ with the same procedure as Algorithm~\ref{alg:entropy_mpi};
%  \STATE // Read its own discovered communities.
%  \STATE Read $C'_p$ with the same procedure as Algorithm~\ref{alg:entropy_mpi};
  \STATE // Use $maxNormedComs[]$ and $maxTComs[]$ to record the max item for each ground truth community shown in Equations~(\ref{eq:fmeasure}) and (\ref{eq:nvd}), respectively.
  \STATE Get $maxNormedComs[]$ and $maxTComs[]$ for each community in $C_p$ with its own $C'_p$;
  \STATE $recvSrc=(p+numProcs-1)\mod{numProcs}$;
  \STATE $sendDst=(p+1)\mod{numProcs}$;
  \STATE // Circulate $C'_p$ in a ring.
  \STATE $receivedMsgNum=0$;
  \WHILE{$receivedMsgNum<(numProcs-1)$}
    \STATE $sendBuf[] \leftarrow C'_p$;
    \STATE $\text{Send}~sendBuf[]~\text{to}~sendDst$;
    \STATE $\text{Receive}~C'_p~\text{from}~recvSrc$ with $recvBuf[]$;
    \STATE Update $maxNormedComs[]$ and $maxTComs[]$ for each community in $C_p$ with received $C'_p$;
    \STATE ++$receivedMsgNum$;
  \ENDWHILE
  \STATE // Use $maxDComs[]$ to record the max item for each detected community shown in Equation~(\ref{eq:nvd}).
  \STATE Get $maxDComs[]$ for each community in $C'_p$ with its own set of ground truth communities $C_p$;
  \STATE // Circulate $C_p$ in a ring.
  \STATE $receivedMsgNum=0$;
  \WHILE{$receivedMsgNum<(numProcs-1)$}
    \STATE $sendBuf[] \leftarrow C_p$;
    \STATE $\text{Send}~sendBuf[]~\text{to}~sendDst$;
    \STATE $\text{Receive}~C_p~\text{from}~recvSrc$ with $recvBuf[]$;
    \STATE Update $maxDComs[]$ for each community in $C'_p$ with received $C_p$;
    \STATE ++$receivedMsgNum$;
  \ENDWHILE
  \STATE Calculate $rankFMeasure$ and also $rankNVD$ with $maxNormedComs[]$, $maxTComs[]$, and $maxDComs[]$ based on Equations~(\ref{eq:fmeasure}) and~(\ref{eq:nvd});
  \STATE Get \textit{F-measure} and $NVD$ by summing $rankFMeasure$ and $rankNVD$ of all $numProcs$ processors;
  \STATE Return \textit{F-measure} and $NVD$;
\end{algorithmic}
\end{algorithm}

We then present the parallel algorithm to calculate the cluster matching based metrics, \textit{F-measure} and $NVD$. From the definitions of \textit{F-measure} and $NVD$ shown in Equation~(\ref{eq:fmeasure}) and Equation~(\ref{eq:nvd}) respectively, we could learn that in order to calculate \textit{F-measure} and $NVD$, we need to determine for each ground truth community the discovered community that has the largest number of common nodes with it. In addition, to calculate $NVD$, we further need to locate for each discovered community the ground truth community that has the largest number of common nodes with it. Hence, in the algorithm, both $C'_p$ and $C_p$ are circulated to each processor in a ``ring''. This ``shifting'' of $C'_p$ and $C_p$ will both occur $N-1$ times for $N$ processors. For its own $C'_p$ and each received $C'_p$, the processor will calculate its local values of \textit{F-measure} and $NVD$ with its own $C_p$. Moreover, for its own $C_p$ and each received $C_p$, the processor will compute its local values of $NVD$ with its own $C'_p$. Finally, the values of \textit{F-measure} and $NVD$ are the sum of the local values of all $N$ processors. Algorithm~\ref{alg:clustering_mpi} shows our parallel MPI algorithm for computing \textit{F-measure} and $NVD$. It takes $C_p$ and $C'_p$ as parameters. %This algorithm does not change the complexity of calculation, thus the complexity is still $O(|V|(|C'|+|C|))$.

\begin{algorithm}
\caption{MPI\_pair\_counting\_metric($C_p(map)$, $C'_p(map)$)}
\label{alg:pair_mpi}
\begin{algorithmic}[1]
  %\STATE // Read its own ground truth communities of this processor and save as map.
%  \STATE Read $C_p(map)$ from ground truth communities;
%  \STATE // Read the community for the nodes in $C_p(map)$ from discovered community structure.
%  \STATE Read $C'_p(map)$ given $C_p(map)$;
  \STATE Count $rankA11$, $rankA10$, $rankA01$, and $rankA00$ using its own $C_p(map)$ and its own $C'_p(map)$;
  \STATE // Circulate $C'_p(map)$ in a ring.
  \STATE $recvSrc=(p+numProcs-1)\mod{numProcs}$;
  \STATE $sendDst=(p+1)\mod{numProcs}$;
  \STATE $receivedMsgNum=0$;
  \WHILE{$receivedMsgNum<(numProcs-1)$}
    \STATE $sendBuf[] \leftarrow C'_p(map)$;
    \STATE $\text{Send}~sendBuf[]~\text{to}~sendDst$;
    \STATE $\text{Receive}~C'_p(map)~\text{from}~recvSrc$ with $recvBuf[]$;
    \STATE Count $rankA01$ and $rankA00$ using its own $C_p(map)$ and received $C'_p(map)$;
    \STATE ++$receivedMsgNum$;
  \ENDWHILE
  \STATE Calculate $rankRI$, $rankARI$, and $rankJI$ with $rankA11$, $rankA10$, $rankA01$, and $rankA00$;
  \STATE Get $RI$, $ARI$, and $JI$ by summing the $rankRI$, $rankARI$, and $rankJI$ of all $numProcs$ processors;
  \STATE Return $RI$, $ARI$, and $JI$;
\end{algorithmic}
\end{algorithm}

We now demonstrate how to calculate the pair counting based metrics, $RI$, $ARI$, and $JI$, in parallel with MPI. To calculate $RI$, $ARI$, and $JI$, each node in the network needs to traverse all the other nodes so as to get $a_{11}$, $a_{10}$, $a_{01}$, and $a_{00}$. Therefore, each processor reads its own ground truth communities and saves as a map with key being the node id and value being the id of the ground truth community to which this node belongs. We denote the map of nodes with their communities from ground truth community structure as $C_p(map)$. This processor also reads the community information for the nodes in $C_p(map)$ from discovered community structure and saves as a map. We denote the map of nodes with their communities from discovered community structure as $C'_p(map)$. $C_p(map)$ and $C'_p(map)$ have the same subset of nodes but with their community information from ground truth community structure and detected community structure, respectively. In our implementation, we use hash indexed map in order to search the community that a node belongs to quickly. Since each node needs to traverse all the other nodes, thus in the algorithm $C'_p(map)$ is circulated to each processor in a ``ring''. This ``shifting'' of $C'_p(map)$ will also occur $N-1$ times for $N$ processors. For each processor, each node in $C_p(map)$ first traverses the other nodes in its own $C'_p(map)$ to count its local values of $a_{11}$, $a_{10}$, $a_{01}$, and $a_{00}$. Then, this node will traverse the nodes in received $C'_p(map)$ to count only $a_{01}$ and $a_{00}$ because this node is in a different ground truth community with the nodes in received $C'_p(map)$. With $a_{11}$, $a_{10}$, $a_{01}$, and $a_{00}$, the processor could get the local values of $RI$, $ARI$, and $JI$ based on Equations~(\ref{eq:ri}), (\ref{eq:ari}), and (\ref{eq:ji}). Finally, the values of $RI$, $ARI$, and $JI$ are the sum of the local values of all $N$ processors. Algorithm~\ref{alg:pair_mpi} shows our parallel MPI algorithm for computing $RI$, $ARI$, and $JI$. It takes $C_p(map)$ and $C'_p(map)$ as parameters. %This algorithm does not change the complexity of calculation either, thus the complexity is still $O(|V|^2)$.

Finally, we illustrate how to calculate the community quality metrics without ground truth community structure, such as modularity and \textit{Modularity Density}, in parallel with MPI. From Section~\ref{sec:CQM_without}, we could observe that in order to calculate the contribution of a community to these metrics, we only need to obtain the number of edges inside it, the number of edges on the boundary of it, the numbers of edges between it and its neighboring communities, and the sizes of it and its neighboring communities. There is no dependency between processors. Hence, there is no need to transfer communities or to transfer any message between processors. In the algorithm, each processor reads its own set of discovered communities $C'_p$ and its local network, and then calculate its own part for these metrics. At last, the values of these metrics are the sum of the local values of all $N$ processors. We will not show the outline of this algorithm here because of its simplicity. %The complexity of this parallel algorithm is $O(|E|+|C'|^2)$.

\begin{algorithm}
\caption{Pthreads\_pair\_counting\_metric($nodes$)}
\label{alg:pair_pthreads}
\begin{algorithmic}[1]
  \STATE // Thread id is denoted as $threadId$. The number of threads used is denoted as $numThreads$.
  \STATE // The number of nodes in the network.
  \STATE $numNodes=nodes.size()$;
  \FOR{$i=0$ to $numNodes$}
    \IF{ $i\mod{numThreads}==threadId$}
      \STATE $iNode=nodes[i]$;
      \STATE // Traverse all the other nodes for $iNode$.
      \FOR{$j=0$ to $numNodes$}
        \STATE $jNode=nodes[j]$;
        \IF{$iNode~\ne jNode$}
          \STATE Count $a11$, $a10$, $a01$, and $a00$ based on the community information of $iNode$ and $jNode$ from ground truth community structure and discovered community structure;
        \ENDIF
      \ENDFOR
    \ENDIF
  \ENDFOR
  \STATE Calculate $RI$, $ARI$, and $JI$ with $a11$, $a10$, $a01$, and $a00$ based on Equations~(\ref{eq:ri}), (\ref{eq:ari}), and (\ref{eq:ji});
  \STATE Return $RI$, $ARI$, and $JI$;
\end{algorithmic}
\end{algorithm}

\subsection{Parallel Algorithms Based on Pthreads}
\vspace{-0.3em}
The parallel Pthreads algorithms for all the metrics, except the ones based on pair counting, assign subsets of ground truth communities and discovered communities, and also local network to each thread using the same approach adopted in the parallel MPI algorithms introduced in Section~\ref{sec:parallel_mpi}. The difference between the parallel Pthreads algorithms and the parallel MPI algorithms is that the ground truth communities, the detected communities, and the network are globally accessable in Pthreads, while they are locally stored in MPI. The cores of the algorithms to calculate these metrics do not change compared with the parallel MPI algorithms, so we will not present their outlines here. %The complexity of these algorithms does not change.

%--------subfigure------------
\begin{figure*}[!t]
\centering
\setlength{\belowcaptionskip}{-1em}
\subfigure[Total running time.]{
\label{mpi_ganxis_mt1:subfig:a}
\includegraphics[scale=0.35]{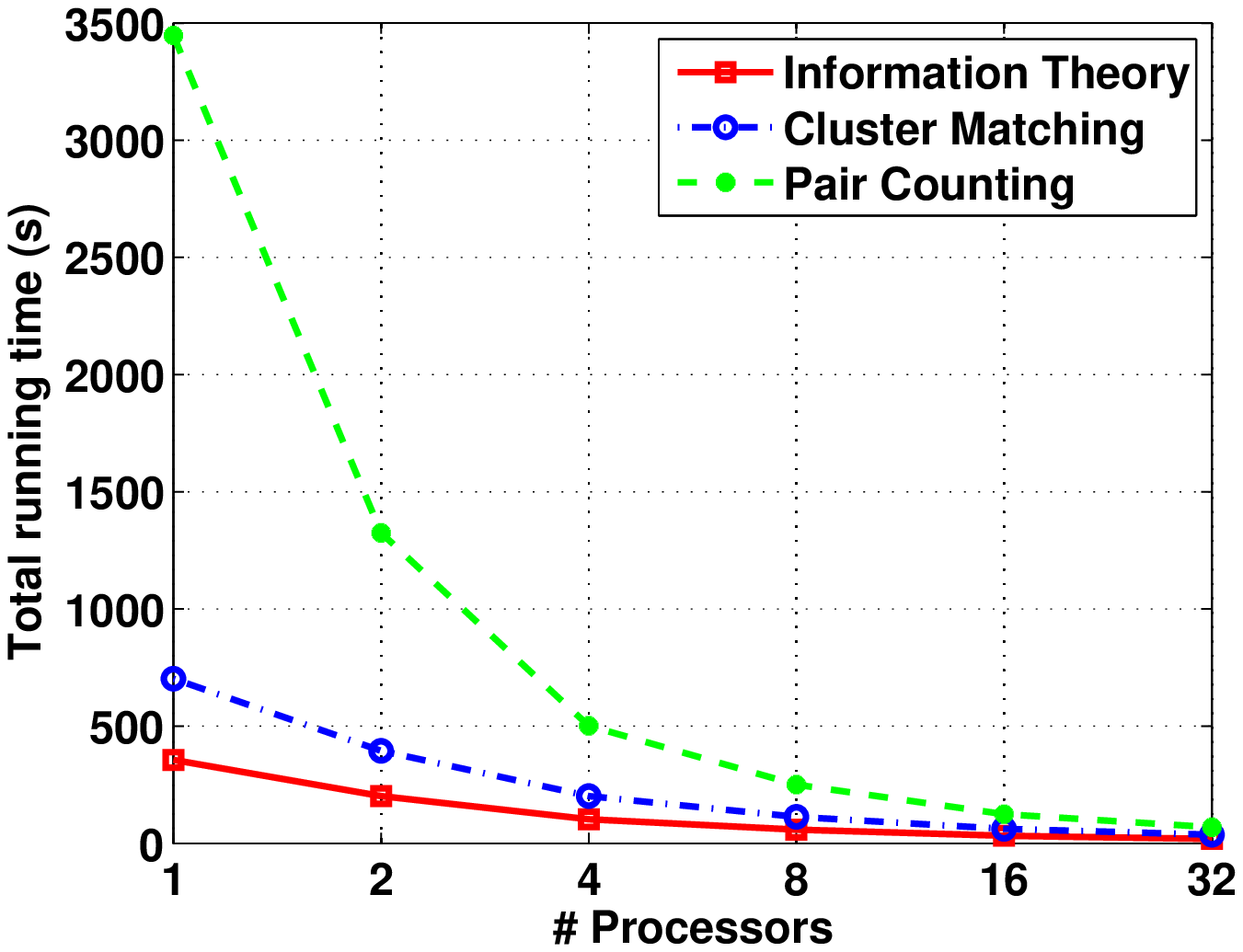}
}
\hspace{0.5em}
\subfigure[Computation time.]{
\label{mpi_ganxis_mt1:subfig:b}
\includegraphics[scale=0.35]{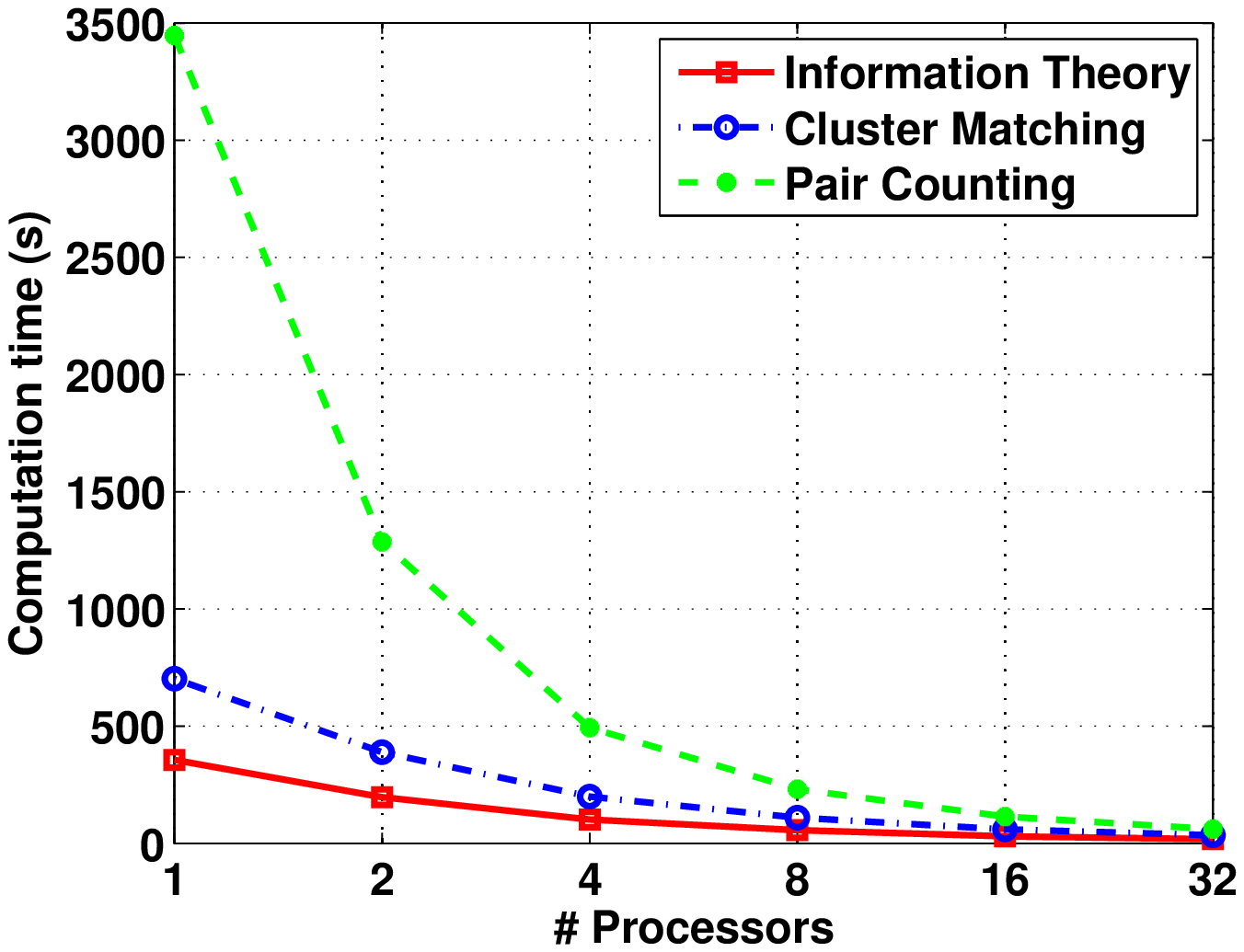}
}
\hspace{0.5em}
\subfigure[Message passing time.]{
\label{mpi_ganxis_mt1:subfig:c}
\includegraphics[scale=0.35]{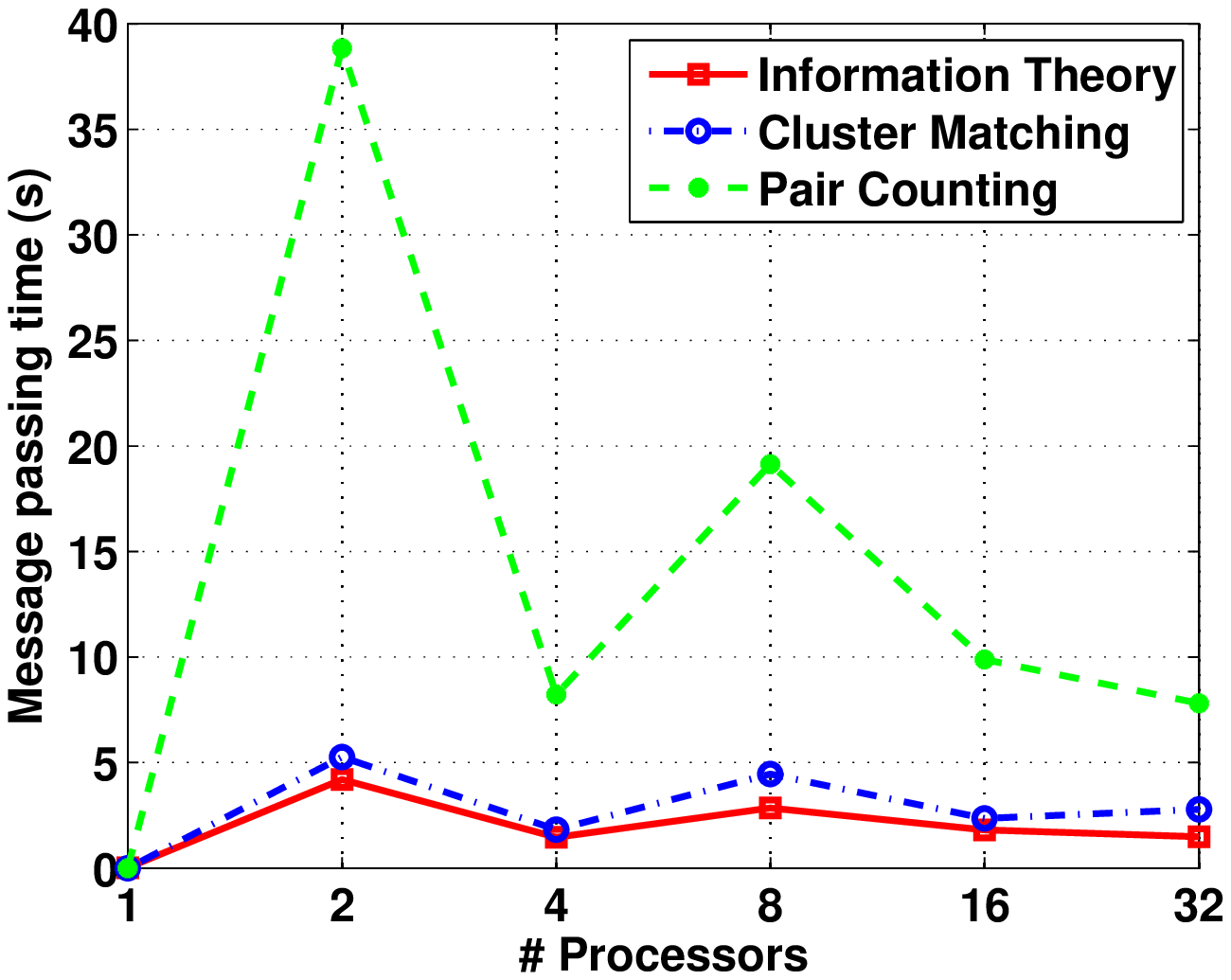}
}
\vspace{-1.2em}
\centering
\caption{The total running time, computation time, and message passing time of the three parallel MPI algorithms for computing the community quality metrics with ground truth community structure on GANXIS.}
\label{mpi_ganxis_mt1}
\vspace{-1.8em}
\end{figure*}

%--------subfigure------------
\begin{figure}[!t]
\centering
\setlength{\belowcaptionskip}{-1em}
\subfigure[Speedup.]{
\label{mpi_ganxis_mt1_measures:subfig:a}
\includegraphics[scale=0.295]{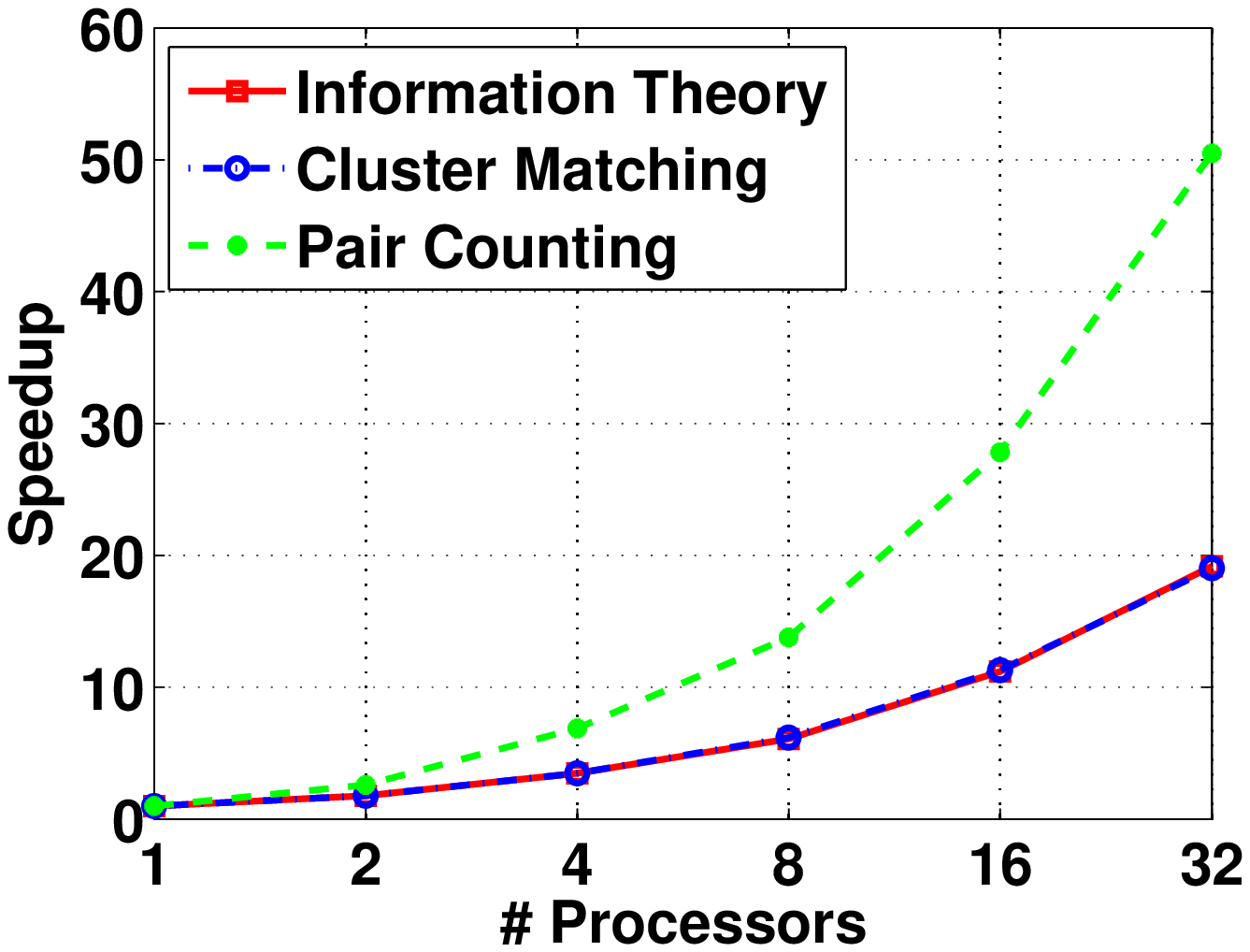}
}
\hspace{-1.7em}
\subfigure[Efficiency.]{
\label{mpi_ganxis_mt1_measures:subfig:b}
\includegraphics[scale=0.295]{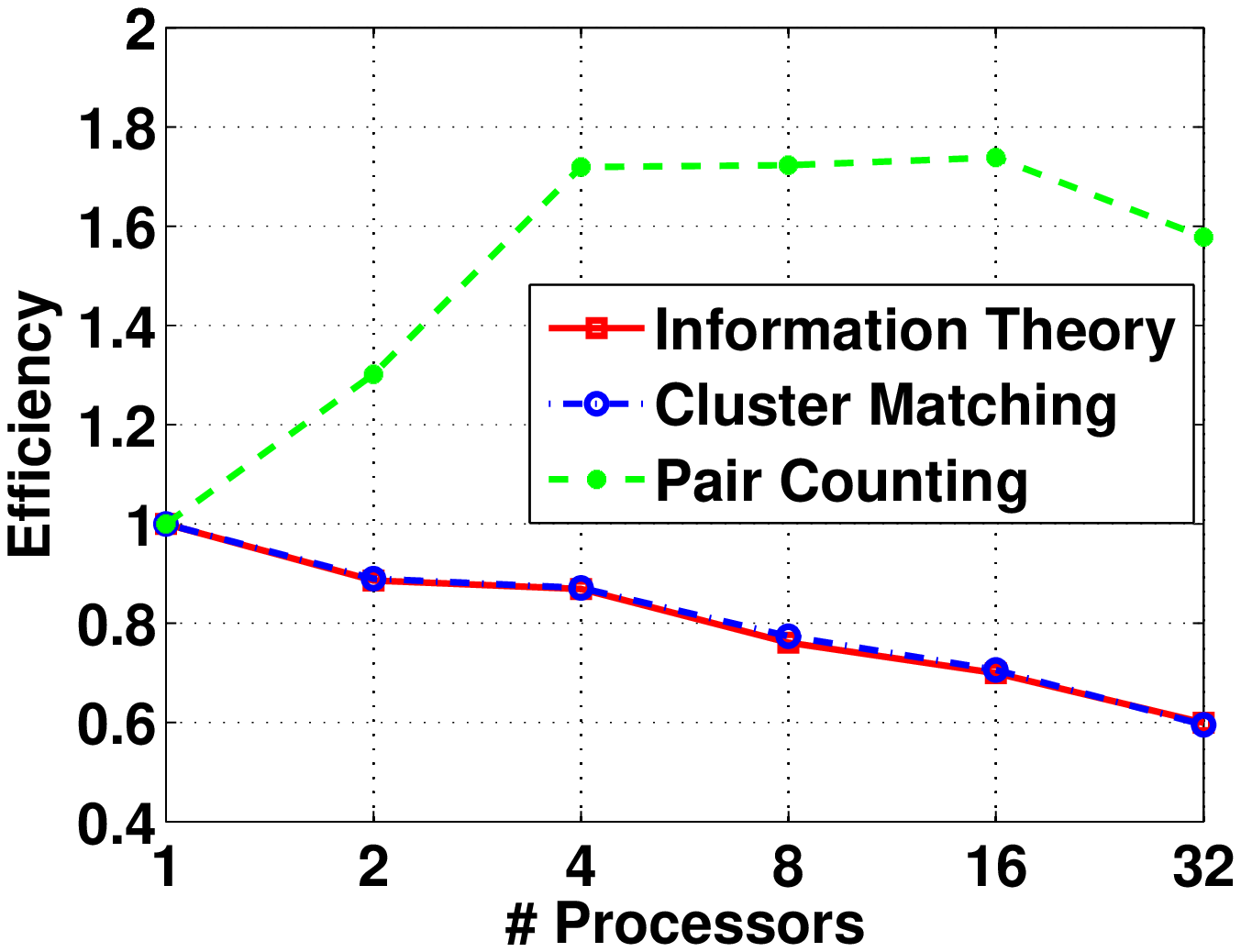}
}
\vspace{-1.7em}
\centering
\caption{The speedup and efficiency of the three parallel MPI algorithms for computing the metrics with ground truth community structure on GANXIS.}
\label{mpi_ganxis_mt1_measures}
\vspace{-1em}
\end{figure}

In the parallel Pthreads algorithm to calculate the pair counting based metrics, the problem is partitioned with the unit of node instead of the unit of community. Then, each node traverses all the other nodes in the network to count $a11$, $a10$, $a01$, and $a00$. The values of $RI$, $ARI$, and $JI$ can be then calculated based on Equations~(\ref{eq:ri}), (\ref{eq:ari}), and (\ref{eq:ji}). The outline of this parallel Pthreads algorithm is shown in Algorithm~\ref{alg:pair_pthreads}. %The complexity of this algorithm is $O(|V|^2)$.

%\vspace{-0.2em}
\section{Evaluation and Analysis}
\vspace{-0.2em}
In this section, we first introduce the parallel architectures on which we perform runs for our parallel algorithms. Then, we introduce the measures that are used to evaluate the performance of these algorithms. We also give an introduction to LFR benchmark network \cite{LFR} for which we calculate the metrics. Finally, we show the performance results of the parallel MPI and Pthreads algorithms. %for calculating the metrics with and without ground truth community structure.

\vspace{-0.2em}
\subsection{Parallel Computing Architectures}
\vspace{-0.3em}
We perform runs on both distributed memory machine, such as Blue Gene/Q, and shared memory machine, like GANXIS. We vary the number of processors used in GANXIS from 1 to 32 and the number of computing nodes (16 cores for each node) used in Blue Gene/Q from 1 to 256.

\subsubsection{GANXIS}
\vspace{-0.2em}
GANXIS a hyper threaded Linux system operating on a Silicon Mechanics Rackform nServ A422.v3 machine (GANXIS.nest.rpi.edu). Processing power was provided by 64 cores organized as four AMD OpteronTM 6272 (2.1 GHz, 16-core, G34, 16 MB L3 Cache) central processing units operating over a shared 512 GB of Random Access Memory (RAM) (32 x 16 GB DDR3-1600 ECC Registered 2R DIMMs) running at 1600 MT/s Max.

\subsubsection{Blue Gene/Q}
\vspace{-0.2em}
The Blue Gene/Q system that we used is stationed at The Computational Center for Nanotechnology Innovations facility at RPI, Troy, NY. %Blue Gene/Q is the third supercomputer design in the Blue Gene series. It has a peak performance 20 Petaflops. The Blue Gene/Q Compute chip is an 18 core chip. The 64-bit PowerPC A2 processor cores are 4-way simultaneously multithreaded, and run at 1.6 GHz. Each processor core has a SIMD Quad-vector double precision floating point unit (IBM QPX). 16 Processor cores are used for computing. The processor cores are linked by a crossbar switch to a 32 MB eDRAM L2 cache, operating at half core speed.

\vspace{-0.2em}
\subsection{Performance Measures}
\vspace{-0.3em}
We calculate \textit{speedup} using Equation~(\ref{eq:speedup})
\begin{equation}
\label{eq:speedup}
%Speedup = \frac{T_1}{T_p}
Speedup = T_1 / T_p
\end{equation}
where $T_1$ is the running time of the sequential program and $T_p$ is the running time of the parallel program when $p$ processors is adopted. We also compute \textit{efficiency} according to Equation~(\ref{eq:efficiency})
\begin{equation}
\label{eq:efficiency}
%Efficiency = \frac{Speedup}{p}
Efficiency = Speedup / p
\end{equation}
where ~$Speedup$ is the actual speedup calculated according to Equation~(\ref{eq:speedup}) and~$p$ is the number of processors.

Notice that for our experimental results on Blue Gene/Q, $p$ is denoted as the number of computing nodes adopted, $T_1$ is the running time of the parallel program when only 1 node adopted, and $T_p$ is the running time of the parallel program when $p$ nodes adopted.

%--------subfigure------------
\begin{figure*}[!t]
\centering
\setlength{\belowcaptionskip}{-1em}
\subfigure[Total running time.]{
\label{mpi_ccni_mt1:subfig:a}
\includegraphics[scale=0.35]{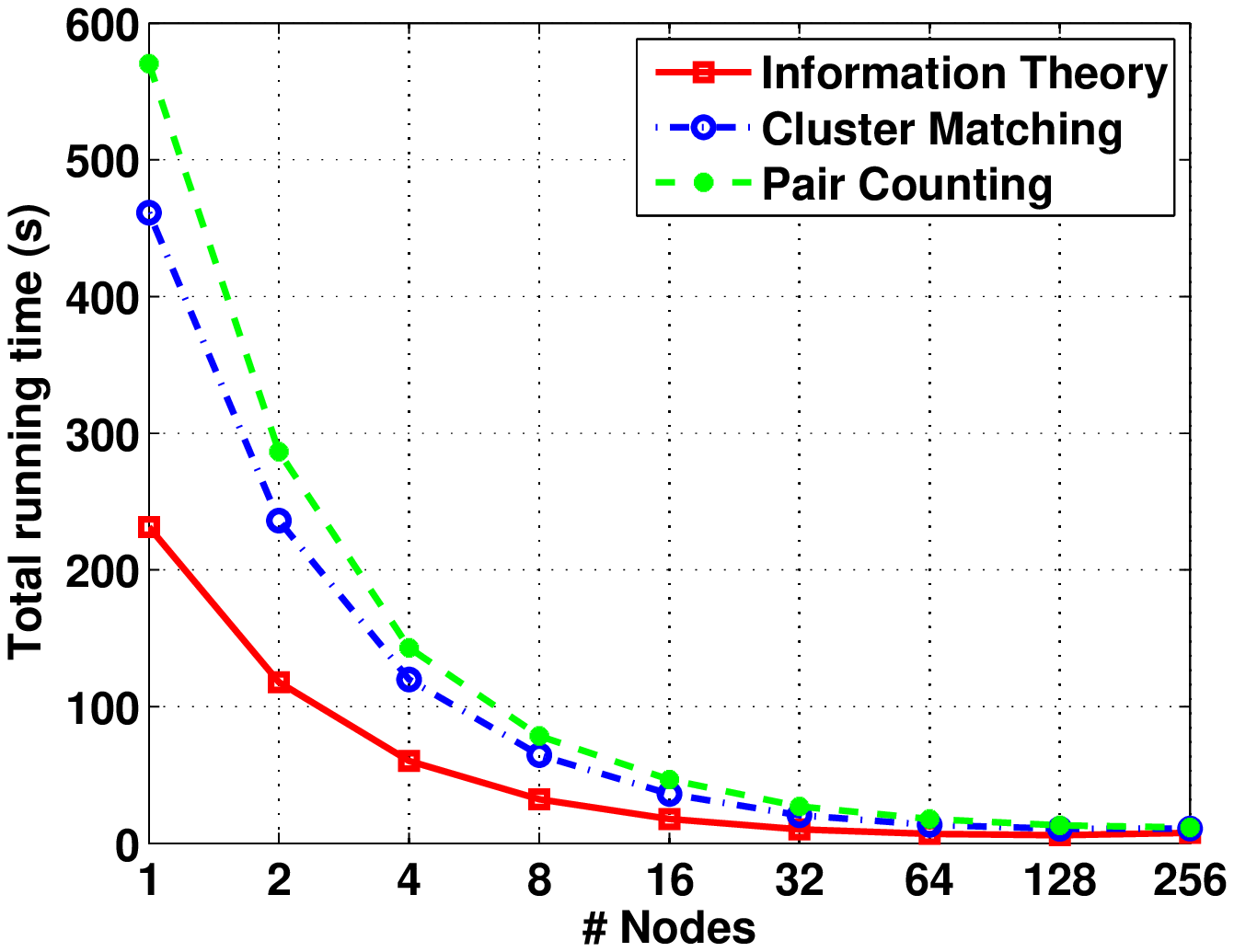}
}
\hspace{0.5em}
\subfigure[Computation time.]{
\label{mpi_ccni_mt1:subfig:b}
\includegraphics[scale=0.35]{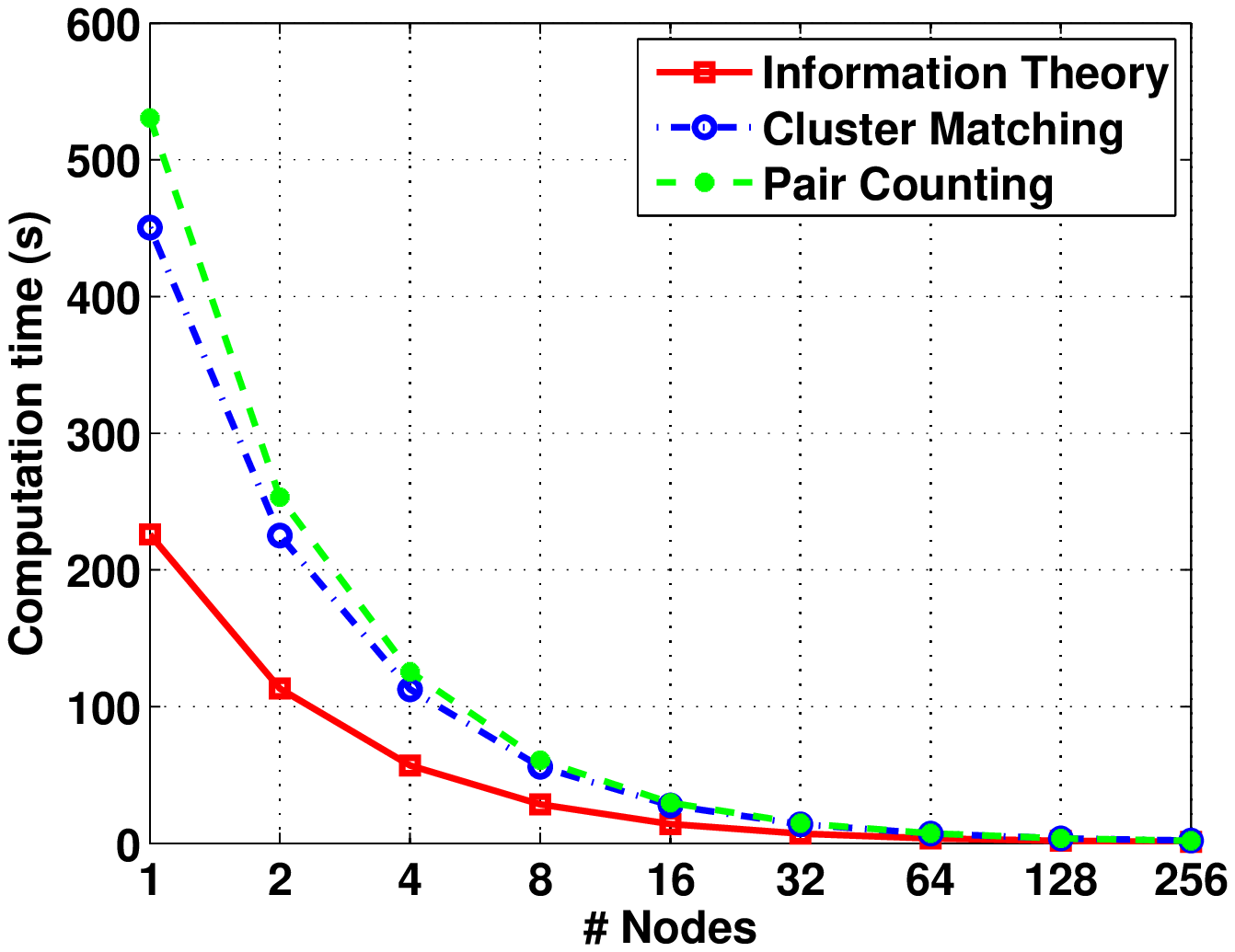}
}
\hspace{0.5em}
\subfigure[Message passing time.]{
\label{mpi_ccni_mt1:subfig:c}
\includegraphics[scale=0.35]{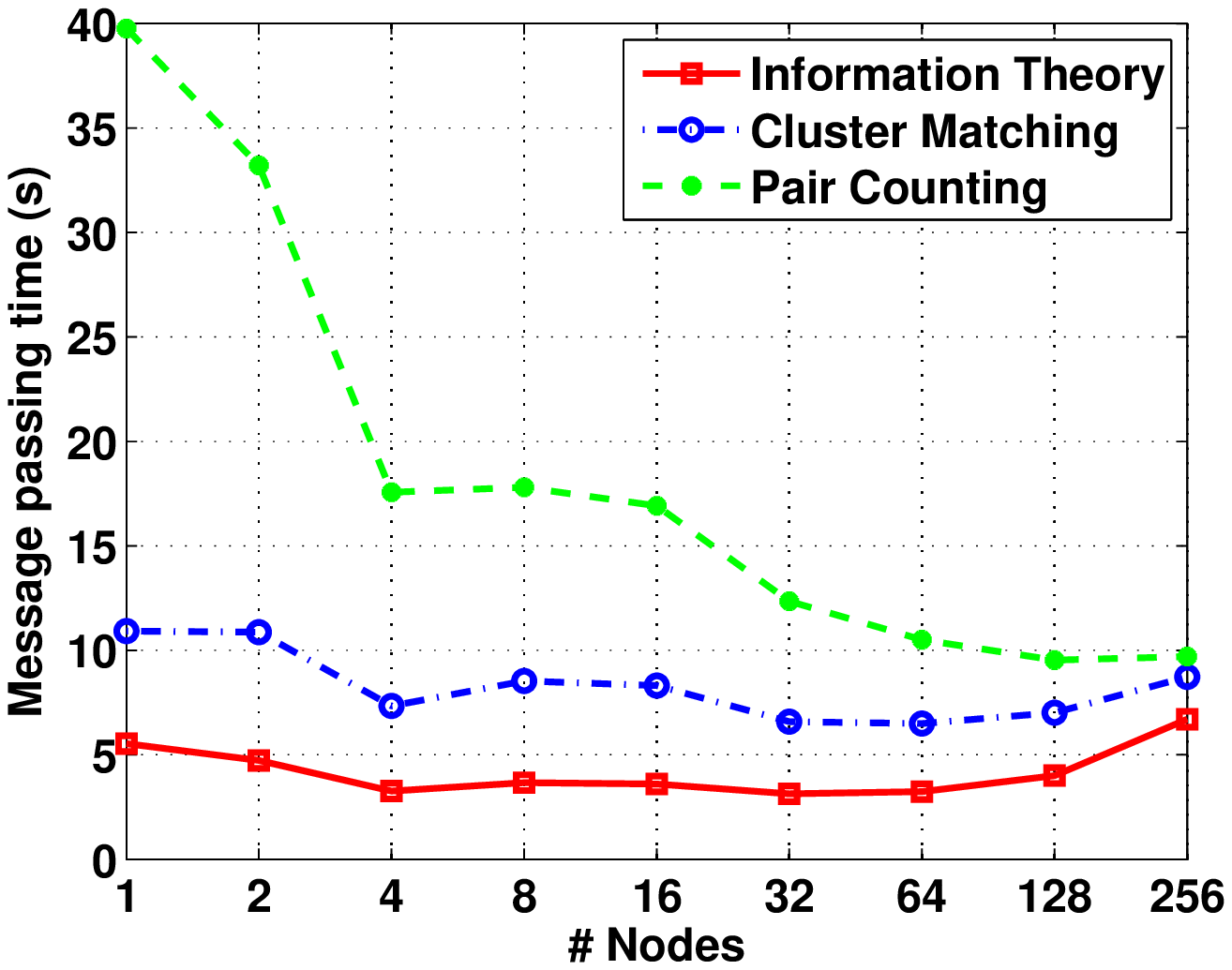}
}
\vspace{-1.2em}
\centering
\caption{The total running time, computation time, and message passing time of the three parallel MPI algorithms for computing the community quality metrics with ground truth community structure on Blue Gene/Q.}
\label{mpi_ccni_mt1}
\vspace{-1.5em}
\end{figure*}

\subsection{LFR Benchmark Network}
\vspace{-0.2em}
We run our parallel MPI and Pthreads programs to calculate the metrics for LFR benchmark networks \cite{LFR} which have known ground truth community structure. The average node degree of the LFR benchmark networks is set to be $15$ and the maximum node degree is set to be $50$. The exponent $\gamma$ for the degree sequence is $2$. The exponent $\beta$ for the community size distribution is $1$. The mixing parameter $\mu$ is equal to $0.3$. %It means that each node shares a fraction $(1- \mu)$ of its edges with the other nodes in its community and shares a fraction $\mu$ of its edges with the nodes outside its community.

The LFR benchmark network for testing the parallel programs for calculating the metrics with ground truth community structure has $100,000$ nodes. The ground truth community structure is given when generating the network. The discovered community structure is obtained by using a community detection algorithm called Speaker listener Label Propagation Algorithm (SLPA) \cite{SLPA2012} with threshold parameter $r=0.5$. SLPA gets disjoint communities when $r=0.5$.

We choose two sizes, ten million of nodes ($10,000,000$) and one hundred million of nodes ($100,000,000$), of LFR network to test the parallel programs to compute the metrics without ground truth communities. We calculate the values of these metrics for the ground truth community structure instead of for the discovered community structure since it takes too long to get the detected communities with SLPA.

%--------subfigure------------
\begin{figure}[!t]
\centering
\setlength{\belowcaptionskip}{-1em}
\subfigure[Speedup.]{
\label{mpi_ccni_mt1_measures:subfig:a}
\includegraphics[scale=0.295]{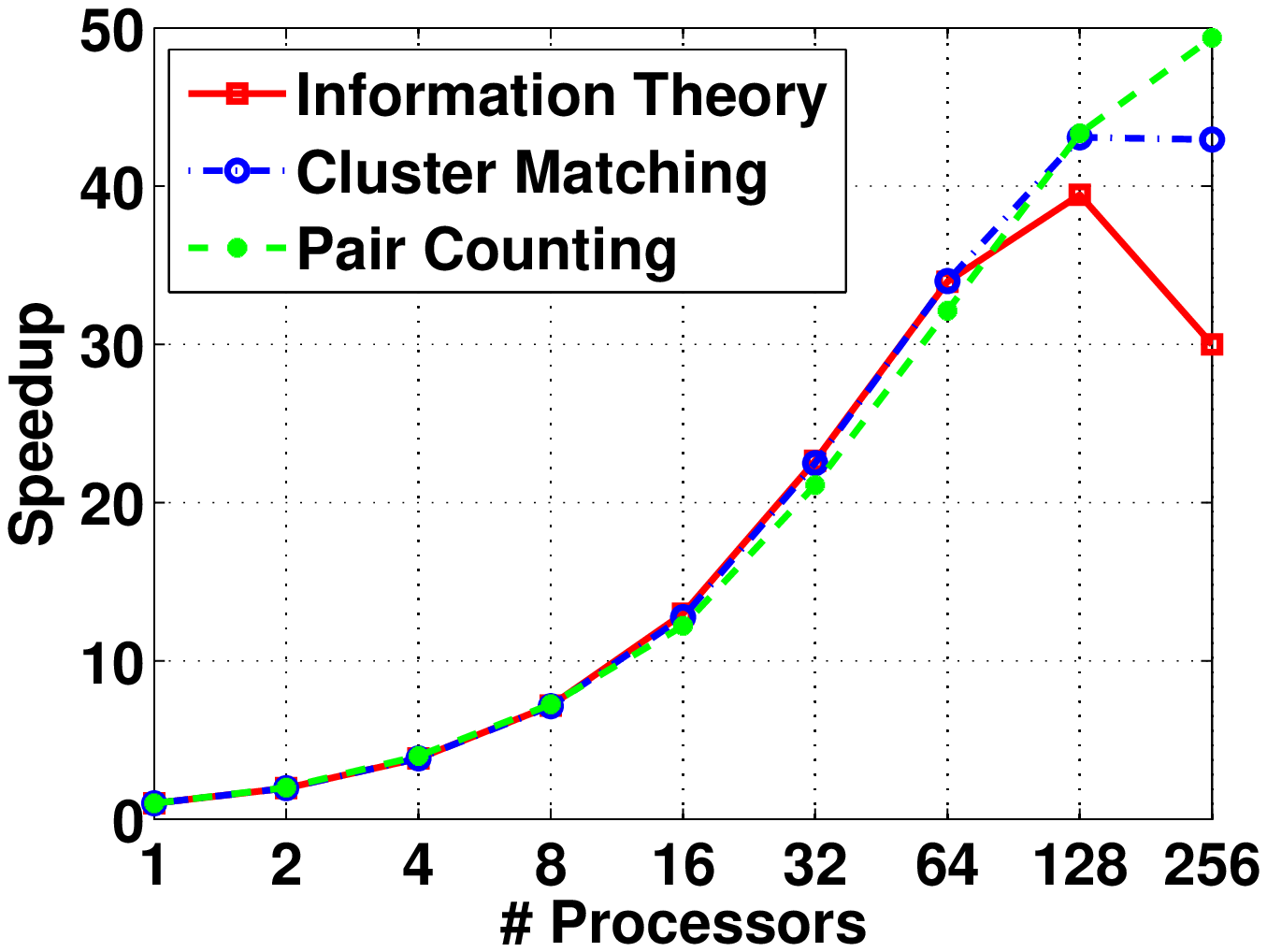}
}
\hspace{-1.7em}
\subfigure[Efficiency.]{
\label{mpi_ccni_mt1_measures:subfig:b}
\includegraphics[scale=0.295]{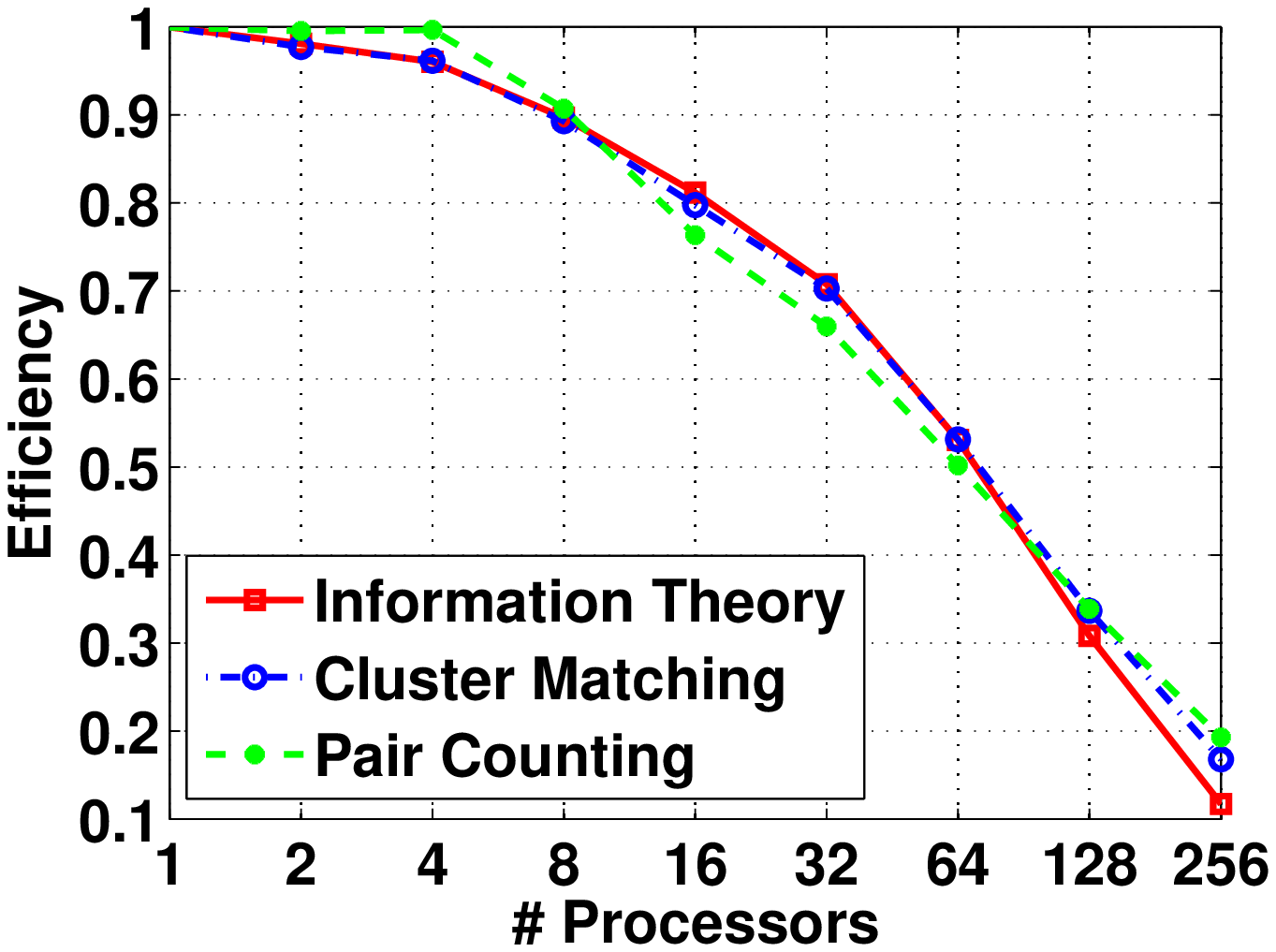}
}
\vspace{-1.7em}
\centering
\caption{The speedup and efficiency of the three parallel MPI algorithms for computing the community quality metrics with ground truth community structure on Blue Gene/Q.}
\label{mpi_ccni_mt1_measures}
\vspace{-0.8em}
\end{figure}

\subsection{Experimental Results}
\vspace{-0.2em}
In this part, we will report the performance results of the parallel MPI and Pthreads algorithms for community quality metrics both with and without ground truth community structure. For the running time, we do not take into account the I/O time. That is, the time of the program to read the ground truth communities, the discovered communities, or the network. Moreover, since GANXIS has 64 processors, every thread in our parallel Pthreads algorithms executes on its dedicated processor. Therefore, threads do not compete for central processing unit (CPU) processors. They execute in parallel, and we can completely ignore thread scheduling issues in our considerations. Because of this we use terms `thread' and `processor' interchangeably when describing the results of the parallel Pthreads algorithms on GANXIS.

\subsubsection{Performance of Parallel Algorithms for Metrics with Ground Truth Community Structure}
\vspace{-0.2em}
Figures~\ref{mpi_ganxis_mt1:subfig:a}, \ref{mpi_ganxis_mt1:subfig:b}, and \ref{mpi_ganxis_mt1:subfig:c} respectively show the total running time, computation time, and message passing time of the three parallel MPI algorithms, Algorithm~\ref{alg:entropy_mpi}, Algorithm~\ref{alg:clustering_mpi}, and Algorithm~\ref{alg:pair_mpi}, to compute the community quality metrics with ground truth communities on GANXIS. Figure~\ref{mpi_ganxis_mt1:subfig:a} indicates that the total running time decreases as the number of processors increases. Figure~\ref{mpi_ganxis_mt1:subfig:b} implies that the computation time decreases as the growth of the number of processors. However, it is shown in Figure~\ref{mpi_ganxis_mt1:subfig:c} that the message passing time goes as a saw shape when the number of processors grows. The saw behavior is the result of GANXIS architecture in which quads of cores uses one shared global memory module. Increasing the number of processors from 1 to 2 introduces the message passing for the first time, adding two more processors helps as the message are placed in the same share memory module for all four cores. In case of 8 processors, some messages start to be moving between two memory modules, which decreases the advantage of having more processors and so on. In addition, Figures~\ref{mpi_ganxis_mt1_measures:subfig:a} and \ref{mpi_ganxis_mt1_measures:subfig:b} present the corresponding speedup and efficiency. From Figure~\ref{mpi_ganxis_mt1_measures:subfig:a}, we can observe that the speedup of all three algorithms grows as the number of processors increases. Figure~\ref{mpi_ganxis_mt1_measures:subfig:b} indicates that the efficiency of Algorithm~\ref{alg:pair_mpi} first increases from 1 to 16 processors and then decreases from 16 to 32 processors, while the efficiency of the other two algorithms always decreases. It is worth noting that the speedup and efficiency of Algorithm~\ref{alg:entropy_mpi} and Algorithm~\ref{alg:clustering_mpi} are almost the same with each other, but both are much smaller than those of Algorithm~\ref{alg:pair_mpi} that is to calculate the pair counting based metrics. The efficiency of Algorithm~\ref{alg:pair_mpi} is even larger than 1, achieving a super-linear speedup. The super-linear speedup is the result of increasing larger cache available on many processors. As the number of processors increases, the volume of data processed on each processor decreases but cache is the same size. Thus, the number of cache misses decreases on each processor, speeding up the execution beyond linear speed up.

Figures~\ref{mpi_ccni_mt1:subfig:a}, \ref{mpi_ccni_mt1:subfig:b}, and \ref{mpi_ccni_mt1:subfig:c} respectively present the total running time, computation time, and message passing time of the three parallel MPI algorithms, Algorithm~\ref{alg:entropy_mpi}, Algorithm~\ref{alg:clustering_mpi}, and Algorithm~\ref{alg:pair_mpi}, for computing the metrics with ground truth communities on Blue Gene/Q. Note that the x-axis is the number of nodes and each node has 16 processors. Thus, the number of processors is the number of nodes times 16. Figure~\ref{mpi_ccni_mt1:subfig:a} demonstrates that the total running time decreases as the number of nodes grows. Figure~\ref{mpi_ccni_mt1:subfig:b} implies that the computation time decreases as the growth of the number of nodes. However, there is no obvious trend of the message passing time. Moreover, Figures~\ref{mpi_ccni_mt1_measures:subfig:a} and \ref{mpi_ccni_mt1_measures:subfig:b} show the corresponding speedup and efficiency. It can be observed from Figure~\ref{mpi_ccni_mt1_measures:subfig:a} that the speedup of the three algorithms grows when the number of nodes increases, except for Algorithm~\ref{alg:entropy_mpi} and Algorithm~\ref{alg:clustering_mpi} when there are 256 nodes, the reason of which is that the message passing time instead of the computation time is the dominant part of the total running time at this case. Figure~\ref{mpi_ccni_mt1_measures:subfig:b} implies that the efficiency of all three algorithms decreases as the growth of the number of nodes.

%--------subfigure------------
\begin{figure*}[!t]
\centering
\setlength{\belowcaptionskip}{-1em}
\subfigure[Total running time.]{
\label{pthreads_ganxis_mt1:subfig:a}
\includegraphics[scale=0.35]{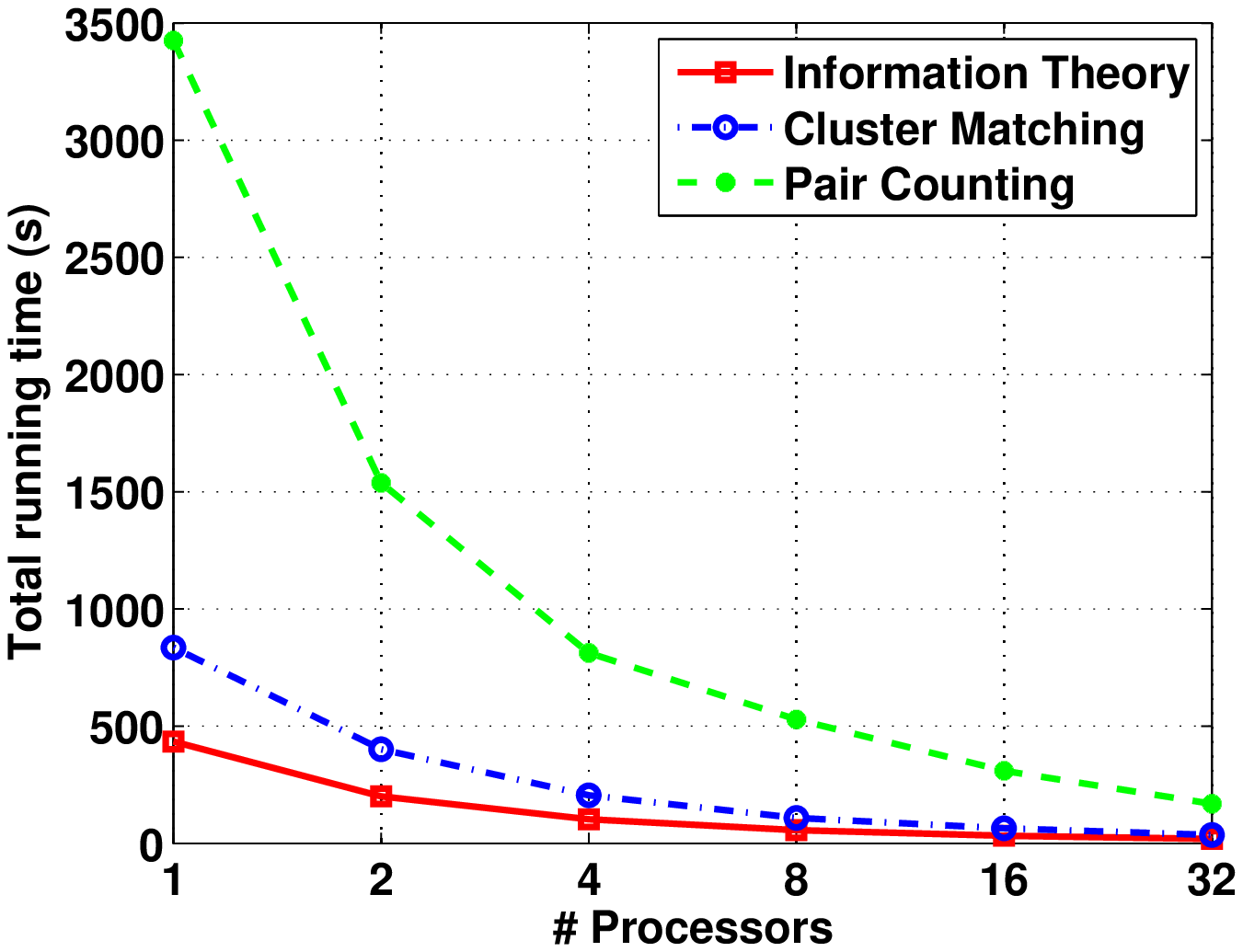}
}
\hspace{0.5em}
\subfigure[Speedup.]{
\label{pthreads_ganxis_mt1:subfig:b}
\includegraphics[scale=0.35]{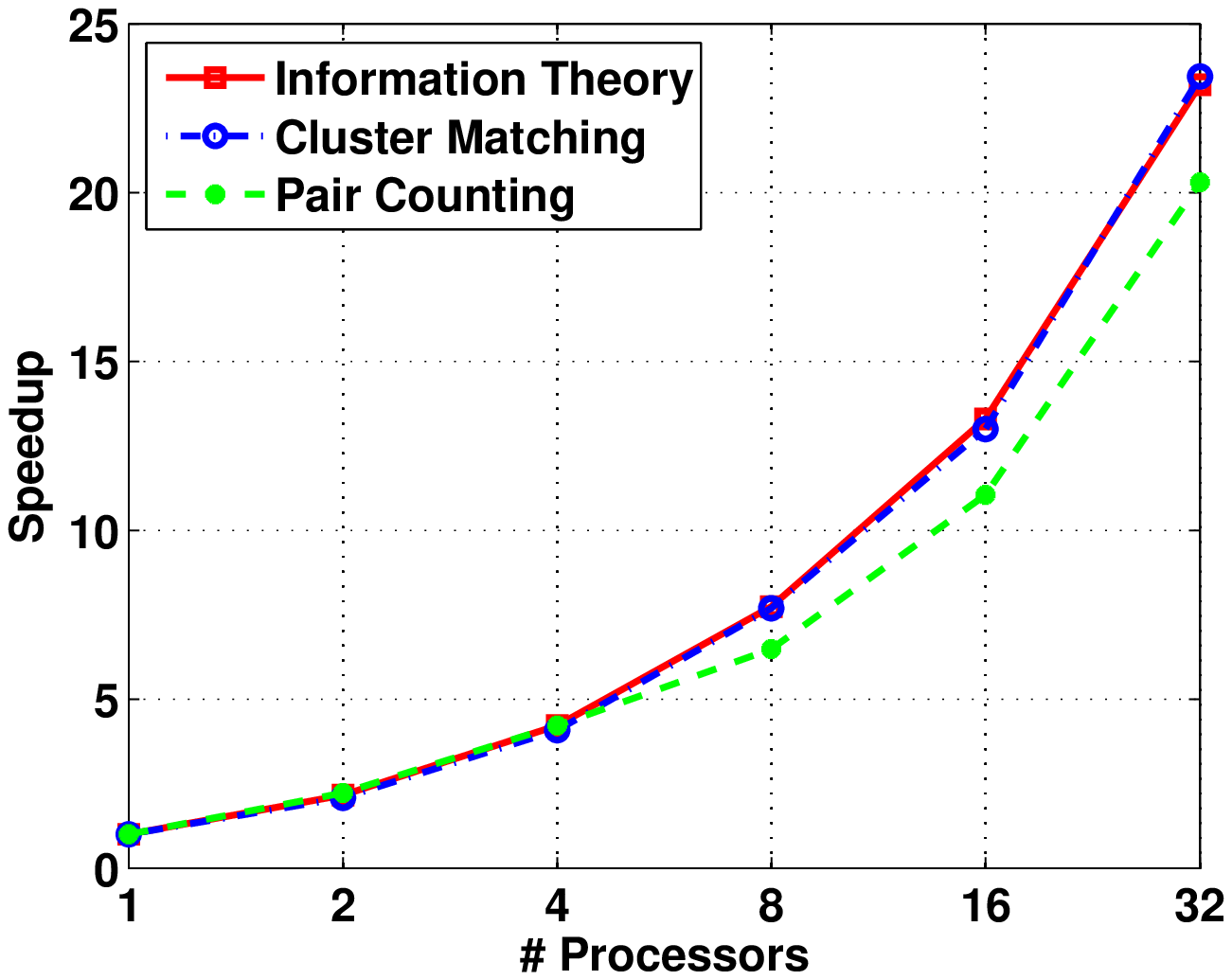}
}
\hspace{0.5em}
\subfigure[Efficiency.]{
\label{pthreads_ganxis_mt1:subfig:c}
\includegraphics[scale=0.35]{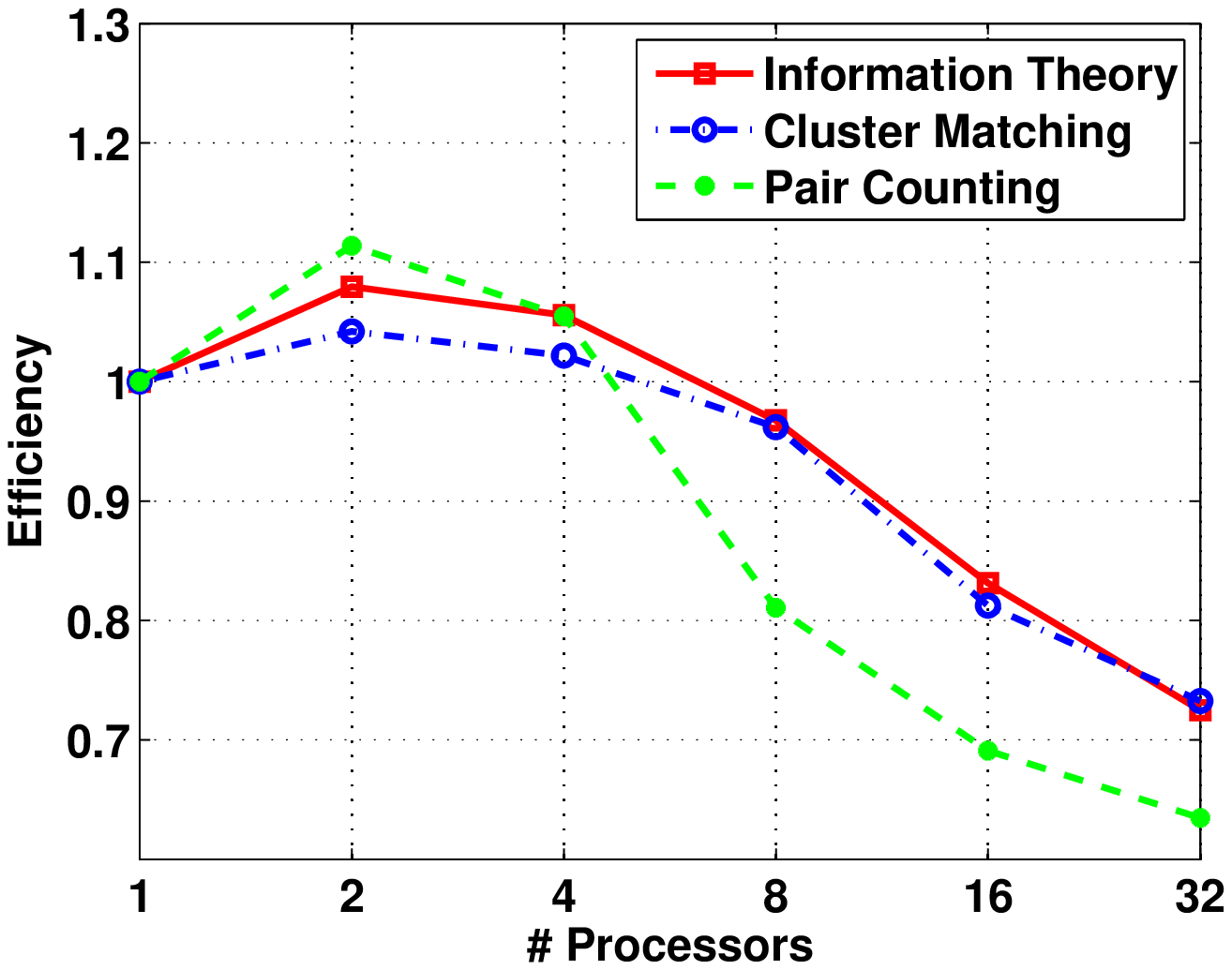}
}
\vspace{-1.2em}
\centering
\caption{The total running time, speedup, and efficiency of the three parallel Pthreads algorithms for computing the community quality metrics with ground truth community structure on GANXIS.}
\label{pthreads_ganxis_mt1}
\vspace{-1.5em}
\end{figure*}

Figures~\ref{pthreads_ganxis_mt1:subfig:a}, \ref{pthreads_ganxis_mt1:subfig:b}, and \ref{pthreads_ganxis_mt1:subfig:c} respectively show the total running time, speedup, and efficiency of the three parallel Pthreads algorithms for computing the metrics with ground truth community structure on GANXIS. Figure~\ref{pthreads_ganxis_mt1:subfig:a} implies that the total running time decreases when the number of processors increases. Figure~\ref{pthreads_ganxis_mt1:subfig:b} indicates that the speedup increases as the growth of the number of processors. However, we could learn from Figure~\ref{pthreads_ganxis_mt1:subfig:c} that the efficiency first grows from 1 to 2 processors and then decreases from 2 to 32 processors. Comparing the total running time in Figure~\ref{pthreads_ganxis_mt1:subfig:a} and in Figure~\ref{mpi_ganxis_mt1:subfig:a}, we could see that the total running time of the three parallel Pthreads algorithms is generally larger than that of the three parallel MPI algorithms. Therefore, we recommend using the parallel MPI algorithms instead of the parallel Pthreads algorithms to calculate the metrics with ground truth community structure on GANXIS (or shared memory machines). Also, it is interesting that the speedup and efficiency of the parallel Pthreads algorithms to calculate the information theory and cluster matching based metrics shown in Figures~\ref{pthreads_ganxis_mt1:subfig:b} and~\ref{pthreads_ganxis_mt1:subfig:c} are larger than those of the corresponding parallel MPI algorithms shown in Figures~\ref{mpi_ganxis_mt1_measures:subfig:a} and~\ref{mpi_ganxis_mt1_measures:subfig:b}. However, the speedup and efficiency of the parallel Pthreads algorithm to calculate the pair counting based metrics shown in Figures~\ref{pthreads_ganxis_mt1:subfig:b} and~\ref{pthreads_ganxis_mt1:subfig:c} are much smaller than those of the corresponding parallel MPI algorithm shown in Figures~\ref{mpi_ganxis_mt1_measures:subfig:a} and~\ref{mpi_ganxis_mt1_measures:subfig:b}. This phenomenon leads to a interesting result that the speedup and efficiency of the parallel Pthreads algorithm to calculate the pair counting based metrics are smaller than those of the other two parallel Pthreads algorithms, while the speedup and efficiency of the parallel MPI algorithm for calculating the pair counting based metrics are much larger than those of the other two parallel MPI algorithms.

%--------subfigure------------
\begin{figure}[!t]
\centering
\setlength{\belowcaptionskip}{-1em}
\subfigure[Total running time.]{
\label{mpi_ganxis_mt0_n70:subfig:a}
\includegraphics[scale=0.2]{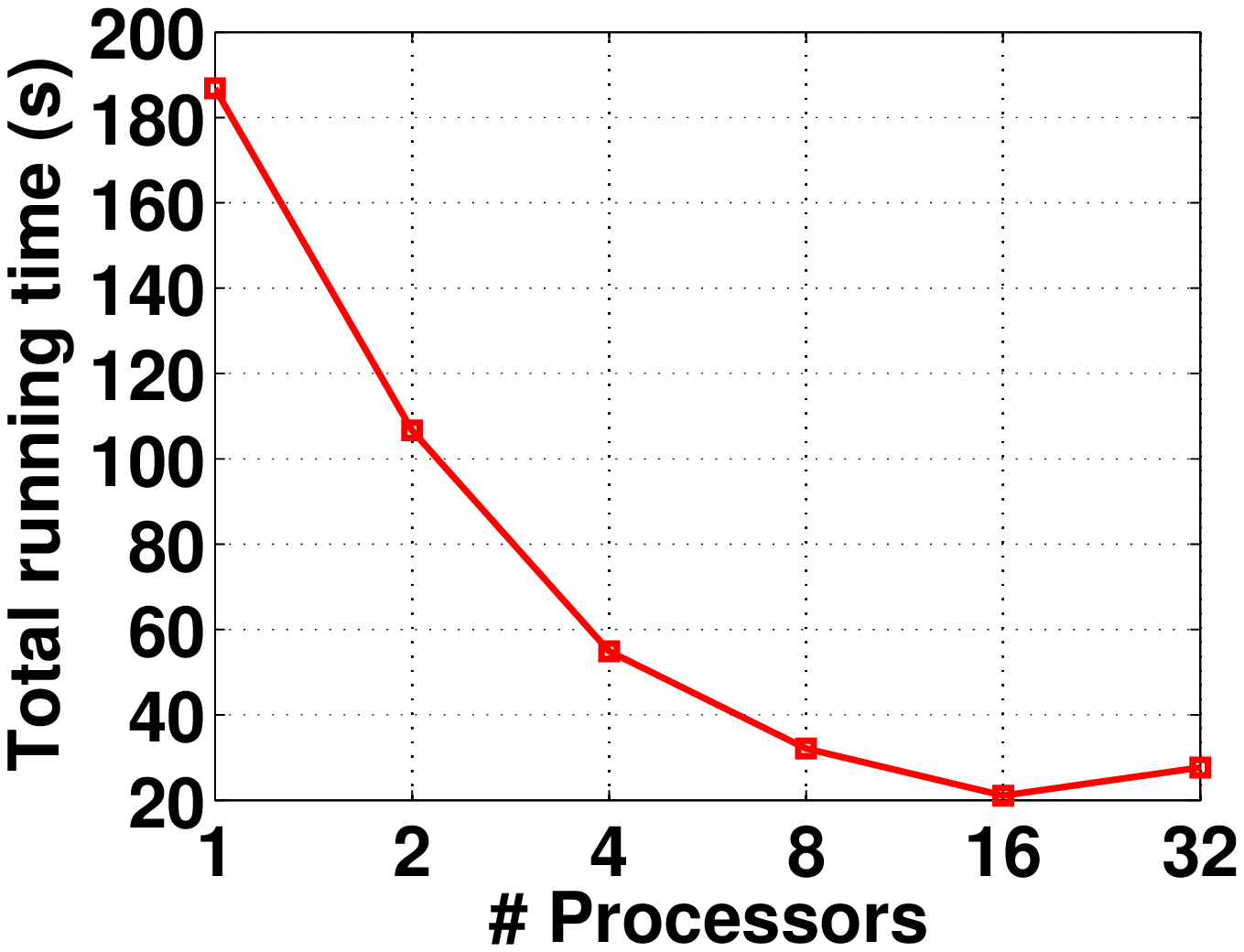}
}
\hspace{-1.8em}
\subfigure[Speedup.]{
\label{mpi_ganxis_mt0_n70:subfig:b}
\includegraphics[scale=0.2]{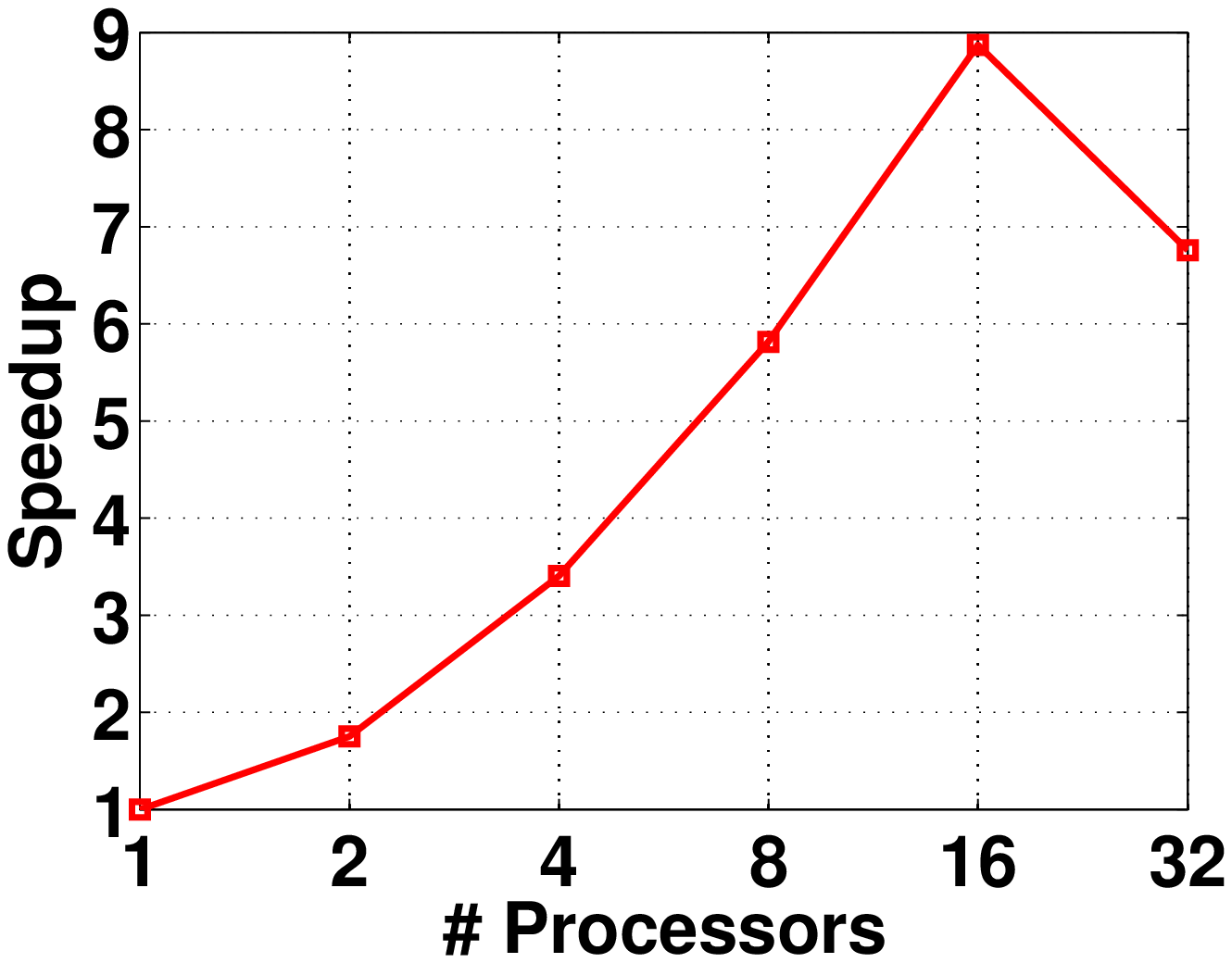}
}
\hspace{-1.8em}
\subfigure[Efficiency.]{
\label{mpi_ganxis_mt0_n70:subfig:c}
\includegraphics[scale=0.2]{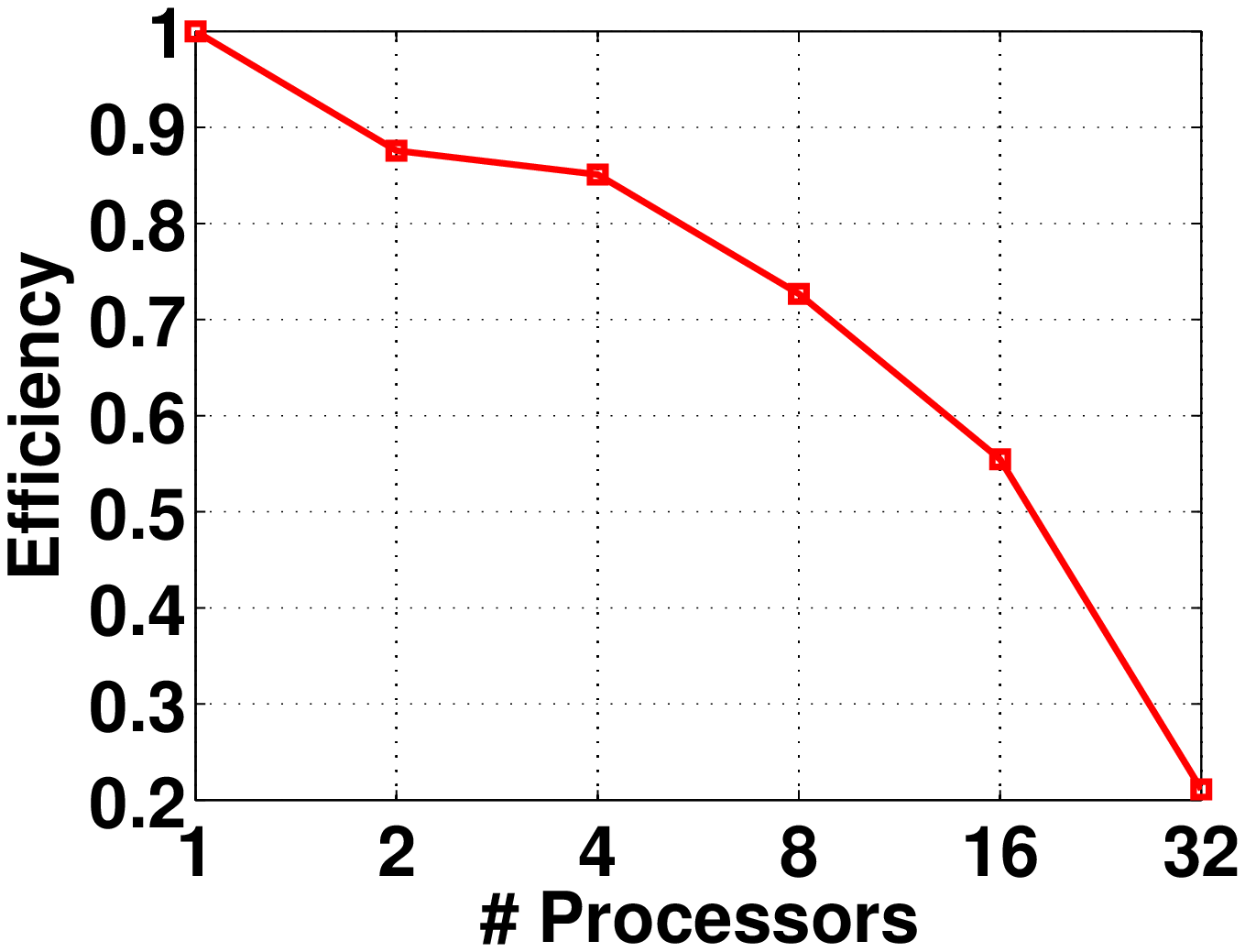}
}
\vspace{-1.7em}
\centering
\caption{The total running time, speedup, and efficiency of the parallel MPI algorithm for computing the community quality metrics without ground truth community structure on GANXIS (The number of nodes of the LFR benchmark network is 10,000,000.).}
\label{mpi_ganxis_mt0_n70}
\vspace{-0.5em}
\end{figure}

%--------subfigure------------
\begin{figure}[!t]
\centering
\setlength{\belowcaptionskip}{-1em}
\subfigure[Total running time.]{
\label{pthreads_ganxis_mt0_n70:subfig:a}
\includegraphics[scale=0.2]{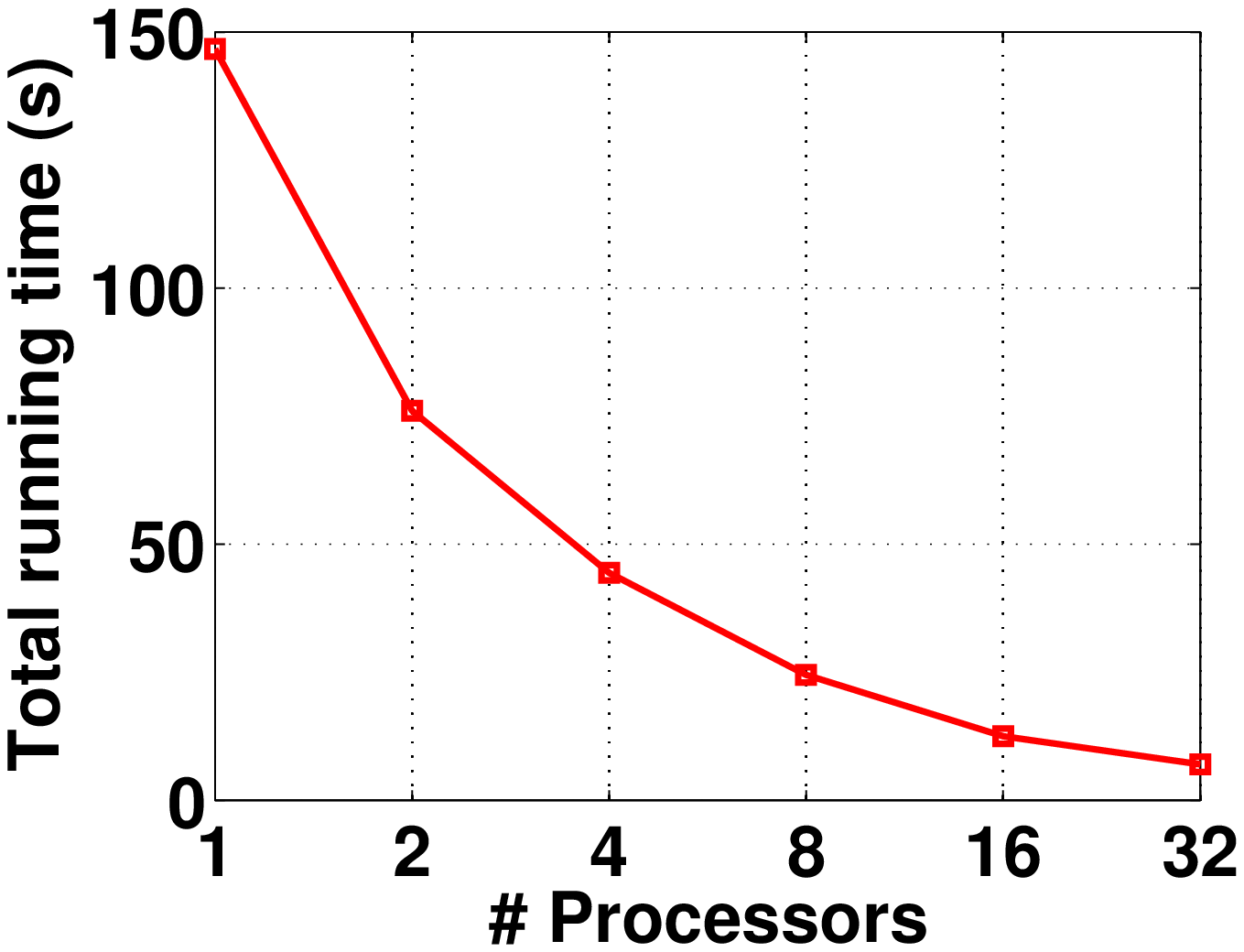}
}
\hspace{-1.8em}
\subfigure[Speedup.]{
\label{pthreads_ganxis_mt0_n70:subfig:b}
\includegraphics[scale=0.2]{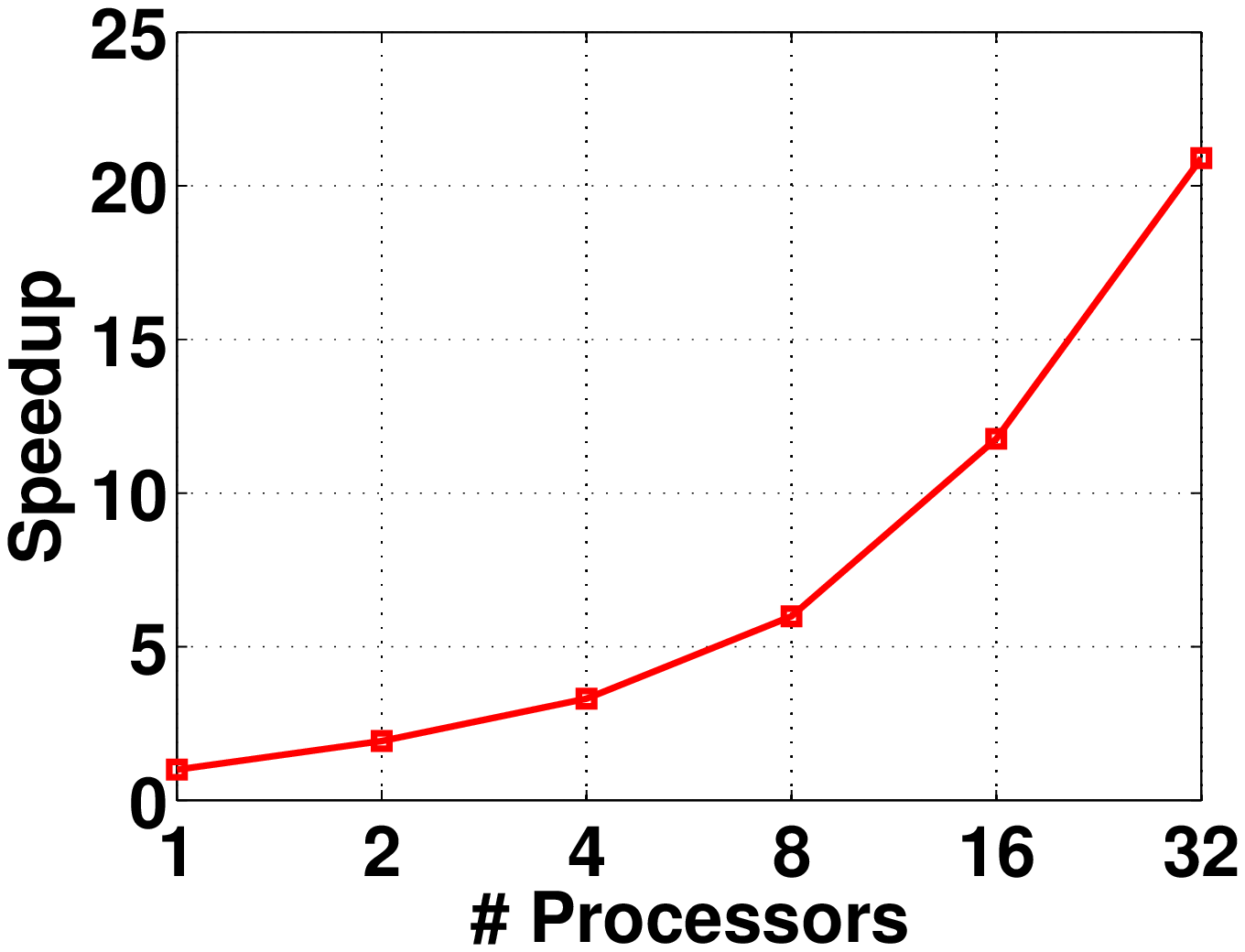}
}
\hspace{-1.8em}
\subfigure[Efficiency.]{
\label{pthreads_ganxis_mt0_n70:subfig:c}
\includegraphics[scale=0.2]{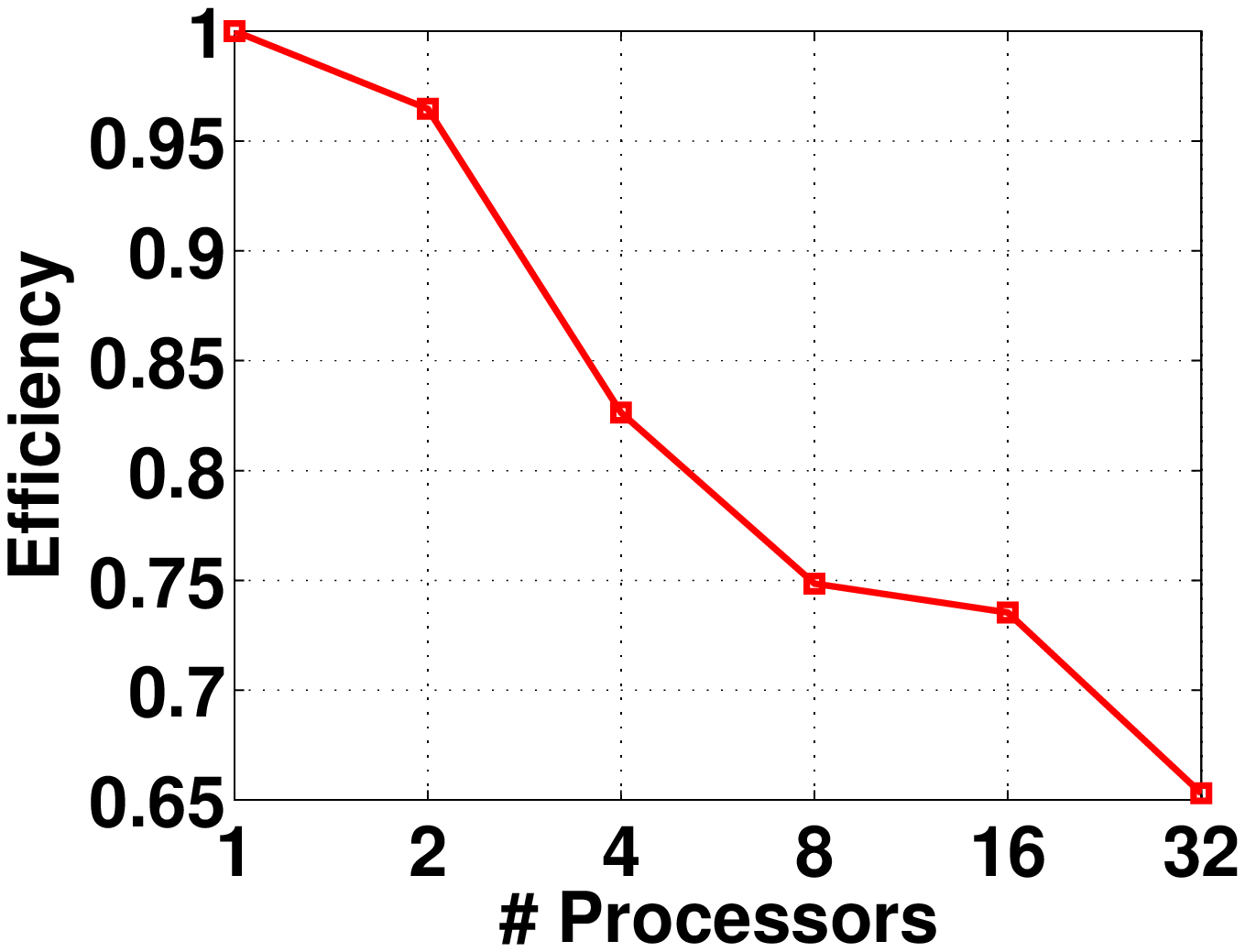}
}
\vspace{-1.7em}
\centering
\caption{The total running time, speedup, and efficiency of the parallel Pthreads algorithm for computing the community quality metrics without ground truth community structure on GANXIS (The number of nodes of the LFR benchmark network is 10,000,000.).}
\label{pthreads_ganxis_mt0_n70}
\vspace{-0.8em}
\end{figure}

\subsubsection{Performance of Parallel Algorithms for Metrics without Ground Truth Community Structure}
\vspace{-0.2em}
Figures~\ref{mpi_ganxis_mt0_n70:subfig:a}, \ref{mpi_ganxis_mt0_n70:subfig:b}, and \ref{mpi_ganxis_mt0_n70:subfig:c} respectively show the total running time, speedup, and efficiency of the parallel MPI algorithm for calculating the metrics without ground truth community structure on GANXIS with the size of the LFR benchmark network being 10,000,000. Figure~\ref{mpi_ganxis_mt0_n70:subfig:a} demonstrates that the total running time first decreases from 1 to 16 processors and then increases a little from 16 to 32 processors. Figure~\ref{mpi_ganxis_mt0_n70:subfig:b} indicates that the speedup first grows from 1 to 16 processors and then decreases from 16 to 32 processors. Thus, we can learn that there is a performance degradation when there are 32 processors. This performance penalty is again caused by the memory banks organization of GANXIS machine. Also, it can be seen from Figure~\ref{mpi_ganxis_mt0_n70:subfig:c} that the efficiency always decreases.

Figures~\ref{pthreads_ganxis_mt0_n70:subfig:a}, \ref{pthreads_ganxis_mt0_n70:subfig:b}, and \ref{pthreads_ganxis_mt0_n70:subfig:c} present the corresponding total running time, speedup, and efficiency of the parallel Pthreads algorithm. We could obverse that the total running time decreases, the speedup grows, and the efficiency decreases as the number of processors increases. Also, the total running time, the speedup, and the efficiency of the parallel Pthreads algorithm are respectively much smaller, larger, and higher than those of the parallel MPI algorithm, compared the results between Figure~\ref{mpi_ganxis_mt0_n70} and Figure~\ref{pthreads_ganxis_mt0_n70}.

\begin{figure}[!t]
\centering
\setlength{\belowcaptionskip}{-1em}
\subfigure[Total running time.]{
\label{mpi_ganxis_mt0_n80:subfig:a}
\includegraphics[scale=0.2]{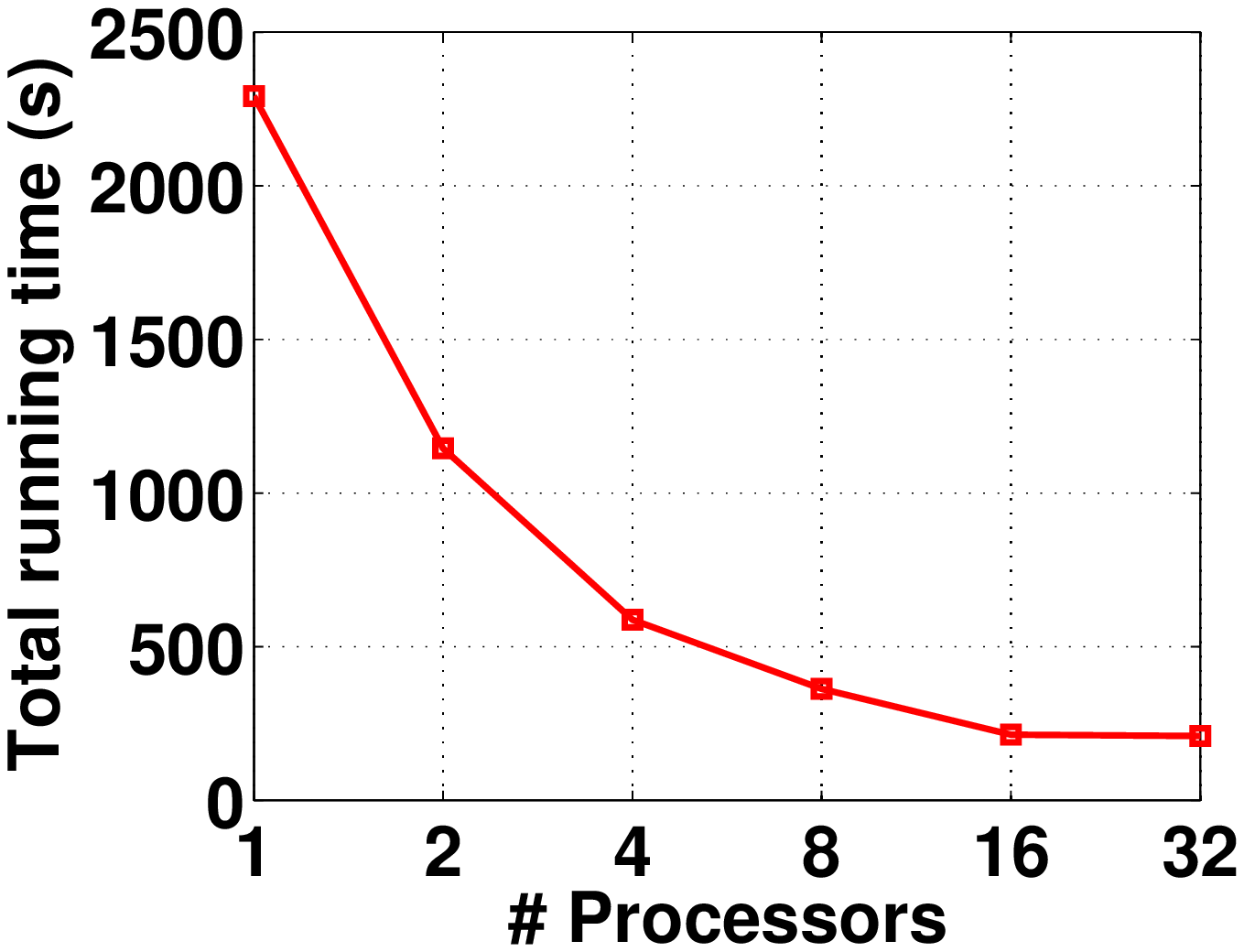}
}
\hspace{-1.8em}
\subfigure[Speedup.]{
\label{mpi_ganxis_mt0_n80:subfig:b}
\includegraphics[scale=0.2]{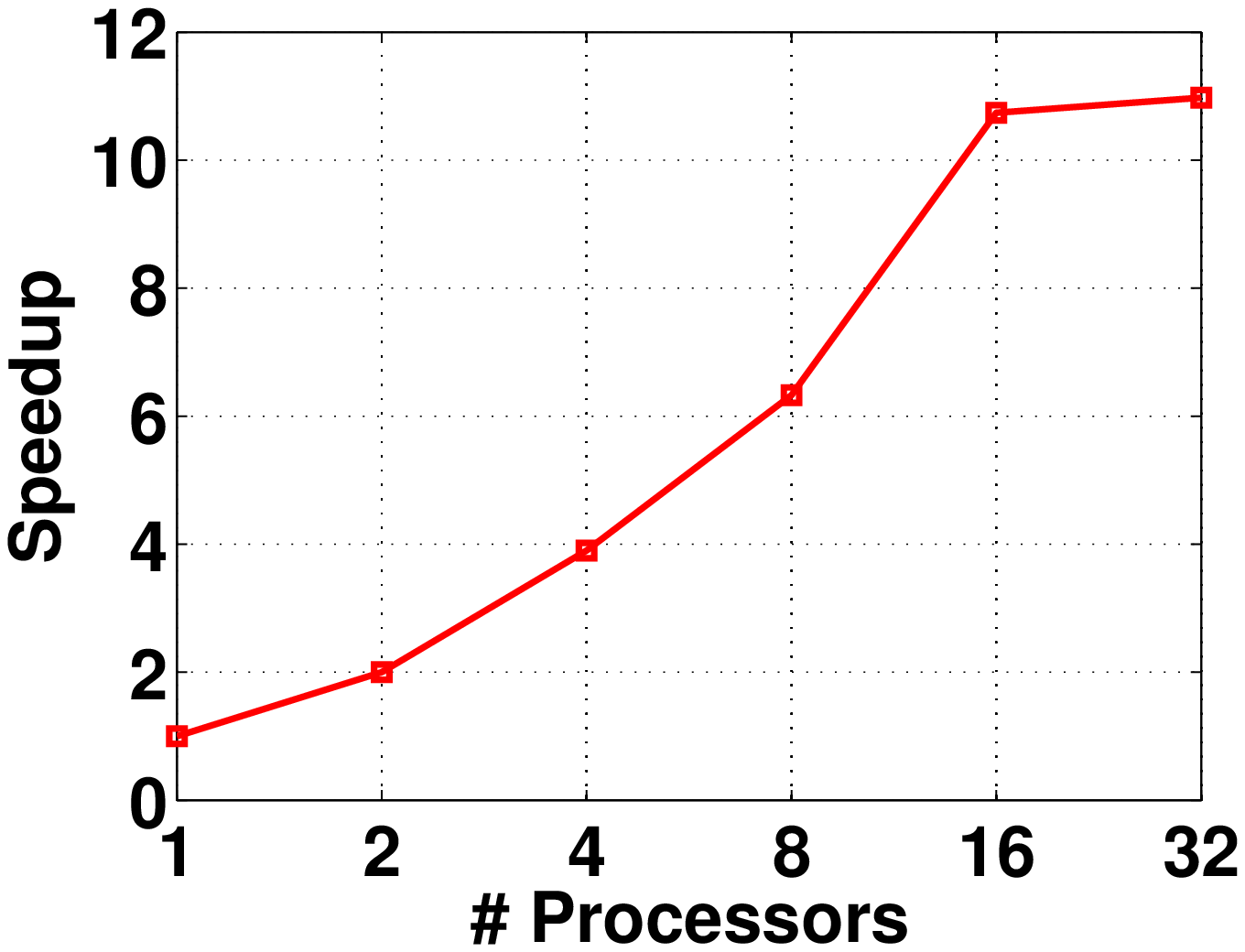}
}
\hspace{-1.8em}
\subfigure[Efficiency.]{
\label{mpi_ganxis_mt0_n80:subfig:c}
\includegraphics[scale=0.2]{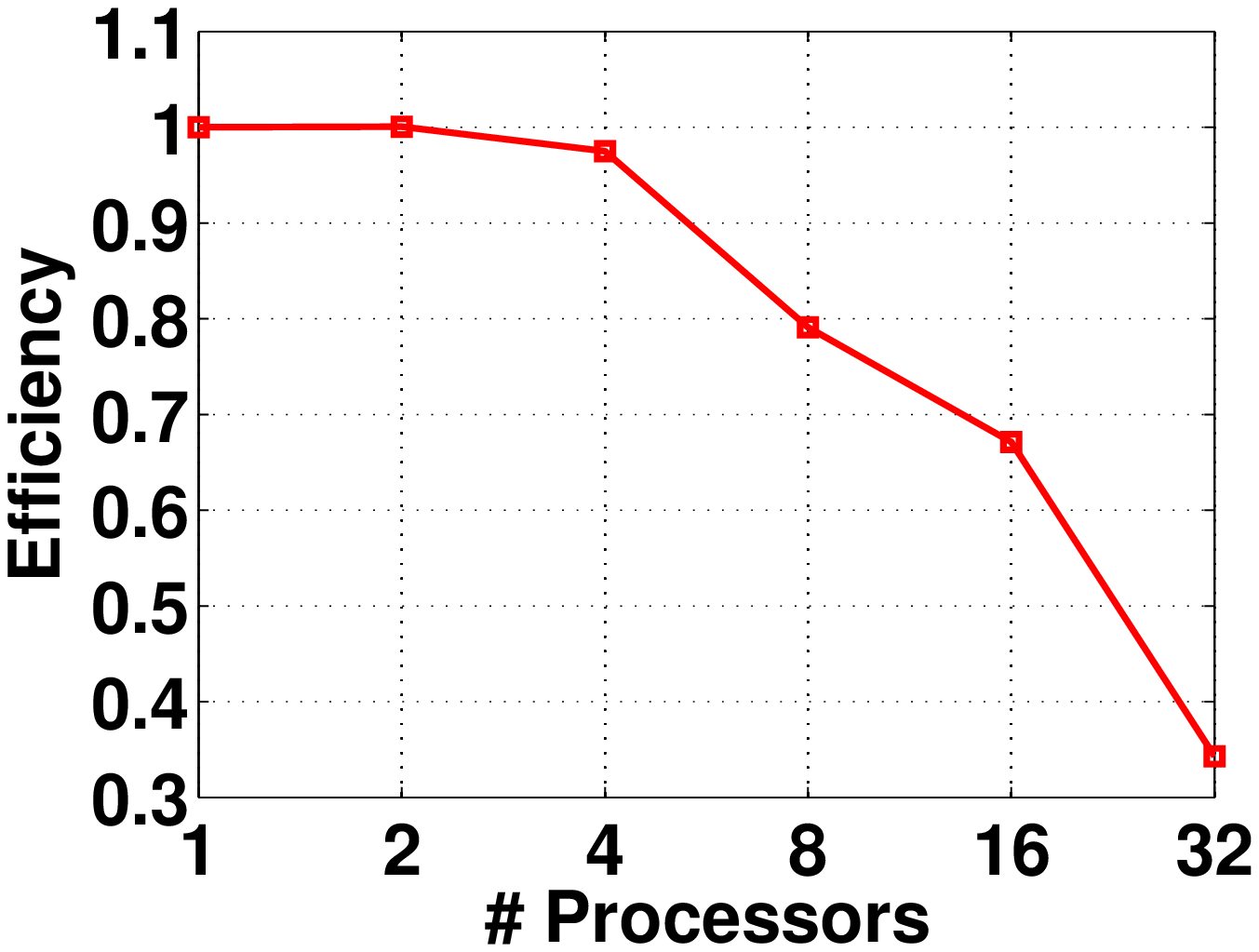}
}
\vspace{-1.7em}
\centering
\caption{The total running time, speedup, and efficiency of the parallel MPI algorithm for computing the community quality metrics without ground truth community structure on GANXIS (The number of nodes of the LFR benchmark network is 100,000,000.).}
\label{mpi_ganxis_mt0_n80}
\vspace{-0.5em}
\end{figure}

Figures~\ref{mpi_ganxis_mt0_n80:subfig:a}, \ref{mpi_ganxis_mt0_n80:subfig:b}, and \ref{mpi_ganxis_mt0_n80:subfig:c} respectively show the total running time, speedup, and efficiency of the parallel MPI algorithm for calculating the metrics without ground truth community structure on GANXIS with the size of the LFR benchmark network being 100,000,000. These three subfigures implies that the total running time decreases, the speedup increases, and the efficiency decreases as the growth of the number of processors. Similarly to the results shown in Figure~\ref{mpi_ganxis_mt0_n70}, Figure~\ref{mpi_ganxis_mt0_n80} demonstrates that there is also a performance degradation when there are 32 processors.

\begin{figure}[!t]
\centering
\setlength{\belowcaptionskip}{-1em}
\subfigure[Total running time.]{
\label{pthreads_ganxis_mt0_n80:subfig:a}
\includegraphics[scale=0.2]{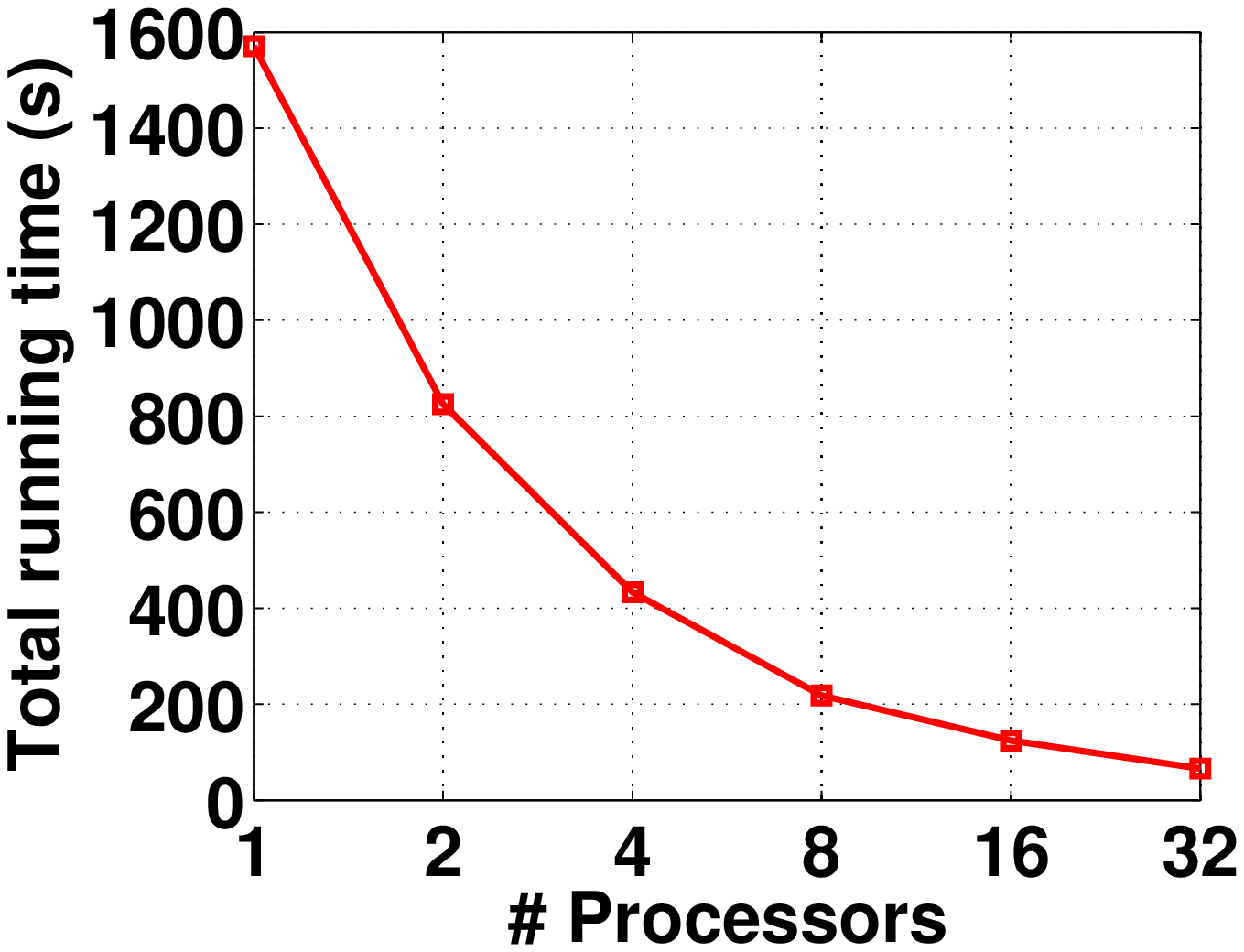}
}
\hspace{-1.8em}
\subfigure[Speedup.]{
\label{pthreads_ganxis_mt0_n80:subfig:b}
\includegraphics[scale=0.2]{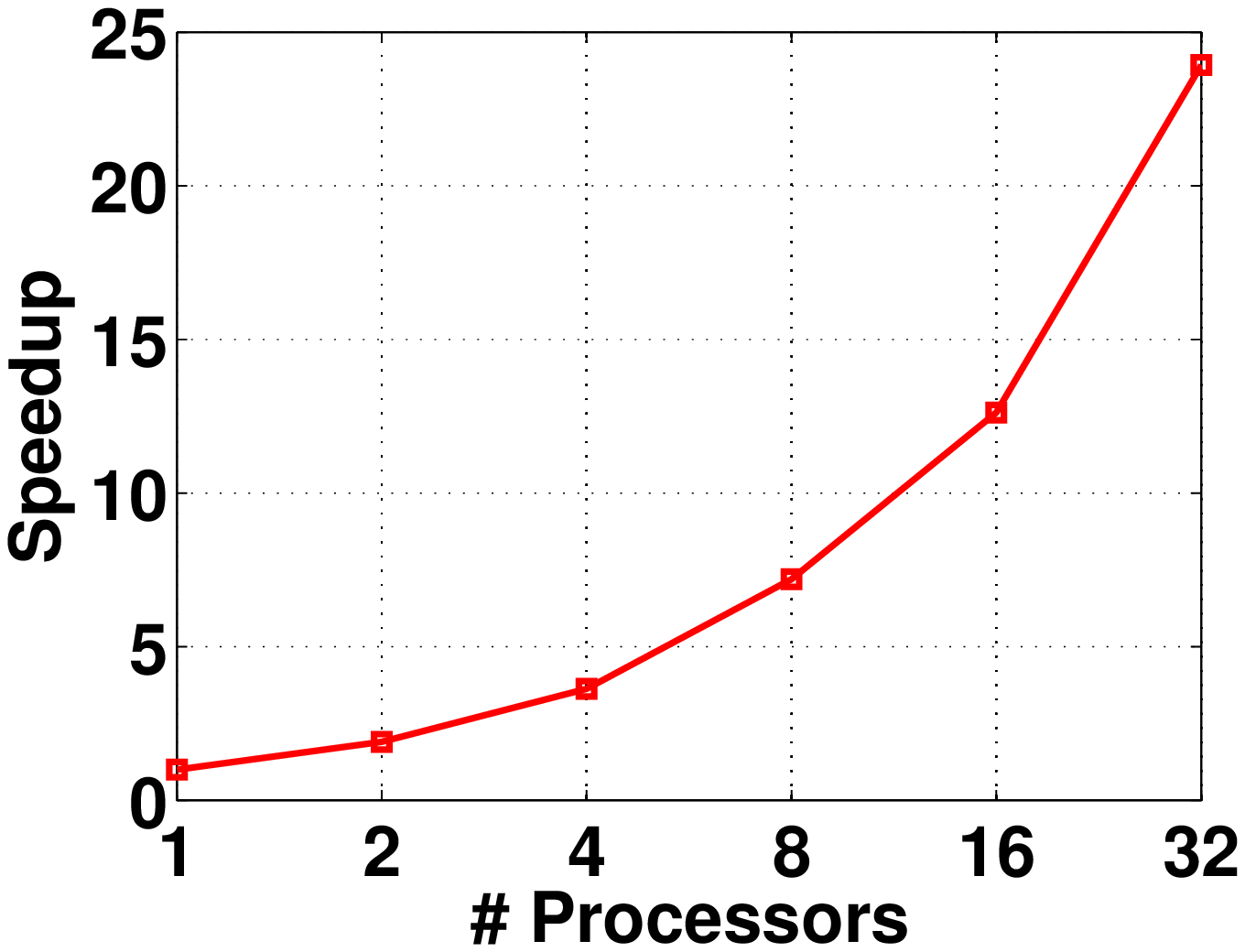}
}
\hspace{-1.8em}
\subfigure[Efficiency.]{
\label{pthreads_ganxis_mt0_n80:subfig:c}
\includegraphics[scale=0.2]{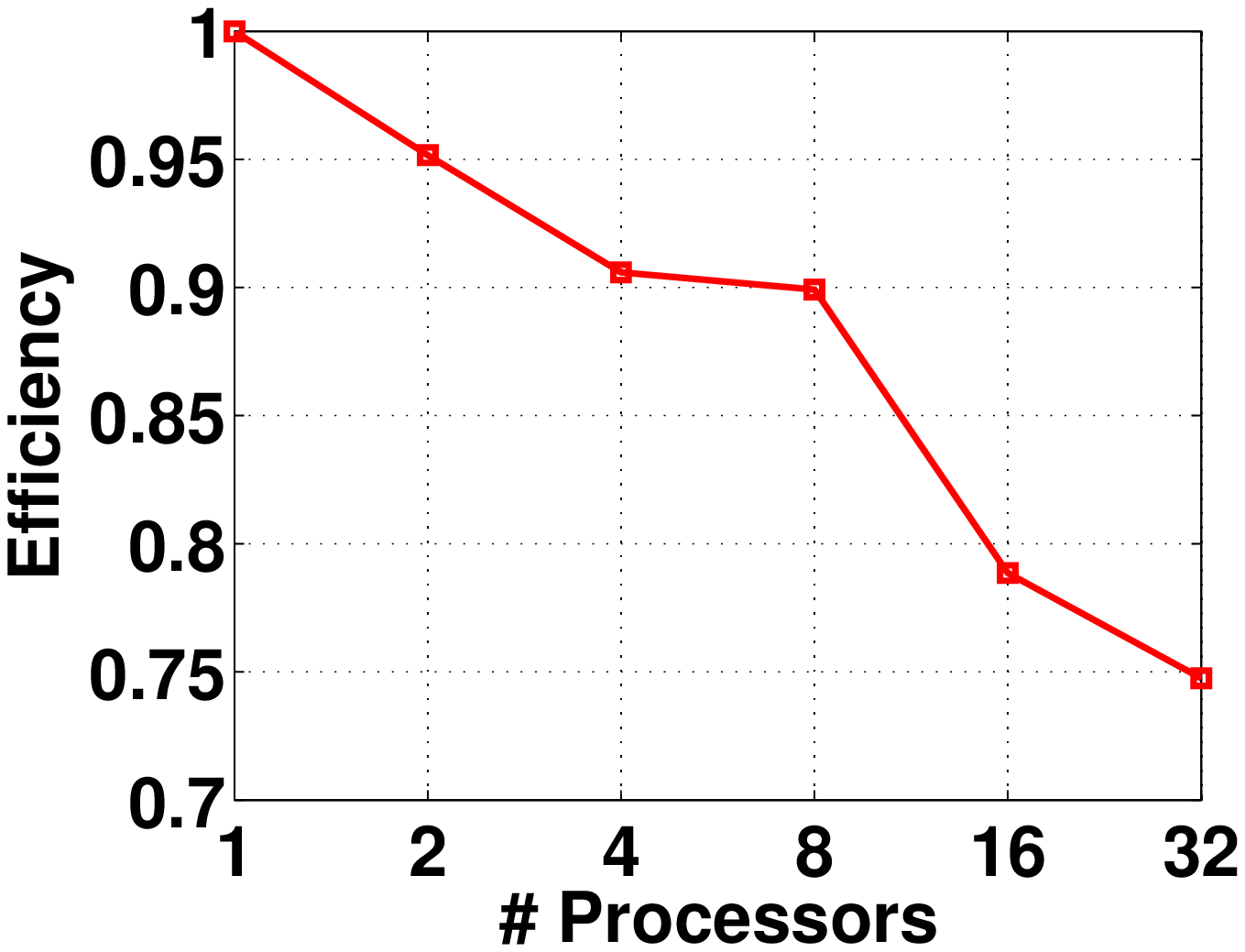}
}
\vspace{-1.7em}
\centering
\caption{The total running time, speedup, and efficiency of the parallel Pthreads algorithm for computing the community quality metrics without ground truth community structure on GANXIS (The number of nodes of the LFR benchmark network is 100,000,000.).}
\label{pthreads_ganxis_mt0_n80}
\vspace{-0.9em}
\end{figure}

Figures~\ref{pthreads_ganxis_mt0_n80:subfig:a}, \ref{pthreads_ganxis_mt0_n80:subfig:b}, and \ref{pthreads_ganxis_mt0_n80:subfig:c} present the corresponding total running time, speedup, and efficiency of the parallel Pthreads algorithm. It shows that the total running time decreases, the speedup grows, and the efficiency decreases when the number of processors increases. Comparing the results between Figure~\ref{mpi_ganxis_mt0_n80} and Figure~\ref{pthreads_ganxis_mt0_n80}, we could observe that the total running time, the speedup, and the efficiency of the parallel Pthreads algorithm are respectively much smaller, larger, and higher than those of the parallel MPI algorithm.

When the size of the LFR benchmark network is 10,000,000, the smallest running time the parallel MPI algorithm achieves is 21.06 seconds at 16 processors, while the smallest running time the parallel Pthreads algorithm achieves is 7.02 seconds at 32 processors. The largest speedup the parallel MPI algorithm gets is 8.87 at 16 processors, while the largest speedup the parallel Pthread algorithm gets is 20.9 at 32 processors. When the size of the LFR benchmark network is 100,000,000, the smallest running time the parallel MPI algorithm can achieve is 208.82 seconds at 32 processors, while the smallest running time the parallel Pthreads algorithm can achieve is 65.66 seconds at 32 processors. The largest speedup the parallel MPI algorithm can get is 10.98 at 32 processors, while the largest speedup the parallel Pthread algorithm can get is 23.92 at 32 processors. Thus, we recommend using the parallel Pthreads algorithm instead of the parallel MPI algorithm to calculate the metrics without ground truth communities on GANXIS (or shared memory machines). There is another point we could get is that the speedup and the efficiency of both parallel MPI algorithm and parallel Pthreads algorithm grow as the size of the network increases.

%--------subfigure------------

%--------subfigure------------

\section{Conclusion}
\vspace{-0.2em}
In this paper, we provide a parallel toolkit, implemented with MPI and Pthreads, to calculate the community quality metrics with and without ground truth community structure. We evaluate their performance on both distributed memory machine, such as Blue Gene/Q, and shared memory machine, for instance GANXIS. We conduct experiments on LFR benchmark networks with the number of nodes being 100,000; 10,000,000; and 100,000,000. The experimental results indicate that both the parallel MPI programs and the parallel Pthreads programs yield a significant performance gain over sequential execution. In addition, we discover that the parallel MPI algorithms perform better than the parallel Pthreads algorithms in terms of total running time, speedup, and efficiency on calculating the metrics with ground truth community structure, while the situation reverses on computing the metrics without ground truth community structure. Therefore, we recommend using the parallel MPI algorithms and the parallel Pthreads algorithm respectively to calculate the metrics with and without ground truth community structure.

%Currently, the parallel toolkit contains only the parallel programs to compute the community quality metrics for disjoint community structure. In the future, we plan to extend the parallel toolkit to calculate also the values of the community quality metrics for overlapping community structure.

% use section* for acknowledgement
\vspace{-0.1em}
\section*{Acknowledgment}
\vspace{-0.3em}
This work was supported in part by the Army Research Laboratory under Cooperative Agreement Number W911NF-09-2-0053 and by the the Office of Naval Research Grant No. N00014-09-1-0607. The views and conclusions contained in this paper are those of the authors and should not be interpreted as representing the official policies either expressed or implied of the Army Research Laboratory or the U.S. Government.

\vspace{-0.1em}
\vspace{-0.3em}
%\bibliography{Parallel_metrics}

% Generated by IEEEtran.bst, version: 1.12 (2007/01/11)

% that's all folks
\end{document}